\documentclass[12pt]{article}
\usepackage{bbm}
\usepackage{stmaryrd}
\usepackage{amsmath}
\usepackage{amsfonts}
\usepackage{latexsym,amssymb,amsmath,mathrsfs}
\newtheorem{theorem}{Theorem}[section]
\newtheorem{lemma}[theorem]{Lemma}
\newtheorem{corollary}[theorem]{Corollary}
\newtheorem{remark}[theorem]{Remark}
\newtheorem{proposition}[theorem]{Proposition}
\newcommand{\mbb}{\mathbb}
\newcommand{\Dlt}{\Delta}
\newcommand{\ptl}{\partial}
\newcommand{\al}{\alpha}
\newcommand{\be}{\beta}
\newcommand{\for}{\mbox{for}}
\newcommand{\vs}{\varsigma}
\newcommand{\stl}{\stackrel}
\newcommand{\vf}{\varphi}
\newcommand{\td}{\tilde}
\newcommand{\sgm}{\sigma}
\newcommand{\es}{\epsilon}
\newcommand{\la}{\langle}
\newcommand{\ra}{\rangle}
\newcommand{\dg}{\dag}

\numberwithin{equation}{section}

\begin{document}

\baselineskip=18pt \begin{center}{\Large \bf Solutions of Navier
Equations and}\end{center}
\begin{center}{\Large \bf Their Representation Structure}\end{center} \vspace{0.2cm}

\begin{center}{\large Bintao Cao}
\end{center}

\begin{center}{Institute of mathematics, Academy of Mathematics \&
Systems Science}\end{center}
\begin{center}{Chinese Academy of Sciences, Beijing 100190, P. R.
China}\end{center}
\begin{center}
Email: caobintao@amss.ac.cn
\end{center}

\vspace{2cm}

 \begin{center}{\Large\bf Abstract}\end{center}

\vspace{1cm} {\small Navier equations are used to describe the
deformation of a homogeneous, isotropic and linear elastic medium in
the absence of body forces. Mathematically, the system is a natural
vector (field) $O(n,\mbb{R})$-invariant generalization of the
classical Laplace equation, which physically describes the vibration
of a string. In this paper, we decompose the space of polynomial
solutions of Navier equations into a direct sum of irreducible
$O(n,\mbb{R})$-submodules and construct an explicit basis for each
irreducible summand. Moreover, we explicitly solve the initial value
problems for Navier equations and their wave-type
extension---Lam\'{e} equations by Fourier expansion and Xu's method
 of solving flag partial differential equations.}

 \vspace{0.8cm}

\section{Introduction}

Classical Laplace equation
$$u_{x_1x_1}+u_{x_2x_2}+\cdots +u_{x_nx_n}=0\eqno(1.1)$$
is one of most fundamental partial differential equations in
mathematics and physics, whose solutions are called {\it harmonic
functions}. Physically it describes the vibration of a string. Its
more general form on Riemmanian manifolds is the main object in
harmonic analysis. A fundamental algebraic characteristic of the
above equation is its invariance under the action of the orthogonal
group $O(n,\mbb{R})$, that is, the space of harmonic functions forms
an $O(n,\mbb{R})$-module. Denote by ${\cal A}_k$ the space of
polynomials of degree $k$ and by ${\cal H}_k$ the space of harmonic
polynomials of degree $k$. The classical harmonic analysis says that
$${\cal A}_k={\cal H}_k\oplus(x_1^2+x_2^2+\cdots+x_n^2){\cal
A}_{k-2}\eqno(1.2)$$ and ${\cal H}_k$ forms an irreducible
$O(n,\mbb{R})$-submodule. By induction, the above conclusion gives a
decomposition of the polynomial algebra into a direct sum of
irreducible $O(n,\mbb{R})$-submodules.  Another closely related
fundamental equation is the wave equation:
$$u_{tt}=u_{x_1x_1}+u_{x_2x_2}+\cdots +u_{x_nx_n}.\eqno(1.3)$$
The solutions of the initial value problems for the Laplace equation
(1.1) and the wave equation (1.3) are elementary known facts
appeared in many textbooks of partial differential equations.

A {\it partial differential equation of flag type} is a linear
differential equation of the form:
$$(d_1+f_1d_2+f_2d_3+\cdots+f_{n-1}d_n)(u)=0,\eqno(1.4)$$
where $d_1,d_2,...,d_n$ are certain commuting locally nilpotent
differential operators on the polynomial algebra
$\mbb{R}[x_1,x_2,...,x_n]$ and $f_1,...,f_{n-1}$ are polynomials
satisfying
$$d_r(f_s)=0\qquad\mbox{if}\;\;r>s.\eqno(1.5)$$
Flag partial differential equations naturally appear in the problem
of decomposing the polynomial algebra (symmetric tensor) over an
irreducible module of a Lie algebra into the direct sum of its
irreducible submodules. Many important linear partial differential
equations in physics and geometry are also of flag type. Xu [X1]
used the grading technique in algebra to develop  methods of solving
such equations. In particular, he found new special functions by
which we are able to explicitly give the solutions of the initial
value problems of a large family of constant-coefficient linear
partial differential equations in terms of their coefficients.

To avoid the confusion with time variable $t$, we use up-index
``$T$" to denote the transpose of a matrix (vector) throughout this
paper. Navier equations
$$\iota_1\Delta (\vec u)+(\iota_1+\iota_2)(\nabla^T\cdot\nabla )(\vec
u)=0\eqno(1.6)$$ are used to describe the deformation of a
homogeneous, isotropic and linear elastic medium in the absence of
body forces (e.g., cf. [G], [M], [T]), where $\vec u$ is an
$n$-dimensional vector-valued function,
$\Dlt=\ptl_{x_1}^2+\ptl_{x_2}^2+\cdots+\ptl_{x_n}^2$ is the Laplace
operator, $\nabla=(\ptl_{x_1},\ptl_{x_2},...,\ptl_{x_n})$ is the
gradient operator, $\iota_1$ and $\iota_2$ are Lam\'{e} constants
 with  $\iota_1>0$, $2\iota_1+\iota_2>0$ and $\iota_1+\iota_2\neq0$. In fact,
 $\nabla^T\cdot\nabla$ is the well-known Hessian operator. Mathematically, the
  above system is a natural vector (field) $O(n,\mbb{R})$-invariant generalization of the
classical Laplace equation. Our first objective in this paper is to
decompose the space of polynomial vectors into a direct sum of
irreducible $O(n,\mbb{R})$-submodules in terms of the irreducible
$O(n,\mbb{R})$-submodules included in the subspace of homogeneous
polynomial solutions of Navier equations.  Unlike the case of the
classical Laplace equation, the subspaces of homogeneous polynomial
solutions of Navier equations are not irreducible.

The wave-type extensions of Navier equations are the well-known
Lam\'{e} equations
$$\vec u_{tt}=\frac{\iota_1}{\iota_1+\iota_2}\Delta(\vec
u)+(\nabla^T\cdot\nabla)(\vec u).\eqno(1.7)$$ Our second objective
in this paper is to solve the initial value problems for Navier
equations and Lam\'{e} equations by Xu's method of solving flag
partial differential equations in [X1] and his
matrix-differential-operator approach in [X2]. Below we give a more
detailed technical introduction.

Olver [O] (1984) found the symmetry group and conservation laws for
Navier equations. Moreover, \"{O}zer [\"{O}] (2003) obtained some
exact solutions of Navier equations with $n=3$ by means of Lie point
transformations. Rodionov [R] (2006)  studied finite solutions of
Lam\'{e} equations by holomorphic expansions. From an algebraic
point of view, it is more convenient to deal with the orthogonal Lie
algebra $o(n,\mbb{R})$ than the orthogonal Lie group $O(n,\mbb{R})$.
They are equivalent under the exponential map from the Lie algebra
to the Lie group in our case.

Similar results as (1.2) hold for some other Lie algebras, too.
Denote by $E_{r,s}$ the square matrix with $1$ as its $(r,s)$-entry
and 0 as the others. Recall the special linear Lie algebra
$$sl(n,\mbb{R})=\sum_{r\neq
s}\mbb{R}E_{r,s}+\sum_{r=1}^{n-1}\mbb{R}(E_{r,r}-E_{r+1,r+1}).\eqno(1.8)$$
Let ${\cal A} $ be the polynomial algebra in
$x_1,...,x_n,y_1,...,y_n.$  Define a representation of
$sl(n,\mbb{R})$ on ${\cal A}$ via
$$E_{r,s}|_{\cal
A}=x_r\ptl_{x_s}-y_s\ptl_{y_r}.\eqno(1.9)$$ Denote by $\mbb{N}$ the
additive semigroup of nonnegative integers. Define
$$x^\al=x_1^{\al_1}x_2^{\al_2}\cdots
x_n^{\al_n}\qquad\for\;\;\al=(\al_1,...,\al_n)\in\mbb{N}^{\:n}.\eqno(1.10)$$
 Set
$${\cal
A}_{\ell_1,\ell_2}=\sum_{\al,\be\in\mbb{N}^{\:n};\;|\al|=\ell_1,\;|\be|=\ell_2}
\mbb{R}x^\al y^\be\qquad\for\;\;\ell_1,\ell_2\in\mbb{N}\eqno(1.11)$$
and define
$${\cal H}_{\ell_1,\ell_2}=\{u\in {\cal
A}_{\ell_1,\ell_2}\mid
u_{x_1y_1}+u_{x_2y_2}+\cdots+u_{x_ny_n}=0\}.\eqno(1.12)$$ Xu [X1]
proved that ${\cal H}_{\ell_1,\ell_2}$ forms an irreducible
$sl(n,\mbb{R})$-submodule and
$${\cal A}_{\ell_1,\ell_2}={\cal H}_{\ell_1,\ell_2}\oplus
(x_1y_1+x_2y_2+\cdots+x_ny_n){\cal
A}_{\ell_1-1,\ell_2-1},\eqno(1.13)$$ where
$x_1y_1+x_2y_2+\cdots+x_ny_n$ is an  $sl(n,\mbb{R})$-invariant. More
importantly, an explicit basis for each ${\cal H}_{\ell_1,\ell_2}$
was constructed in [X1].

Now we assume that ${\cal A}$ is the polynomial algebra in
$x_1,...,x_7$. There exists an action of the simple Lie algebra
${\cal G}^{G_2}$ of type $G_2$ on ${\cal A}$, which keeps
$x_1^2+x_2x_5+x_3x_6+x_4x_7$ invariant (e.g., cf. [H] and [X1]).
Again we denote by ${\cal A}_k$ the subspace of polynomials of
degree $k$ in ${\cal A}$ and define
$$\tilde{\cal H}_k=\{u\in{\cal A}_k\mid
u_{x_1x_1}+u_{x_2x_5}+u_{x_3x_6}+u_{x_4x_7}=0\}.\eqno(1.14)$$ Xu
[X1] proved that $\tilde{\cal H}_k$ forms an irreducible ${\cal
G}^{G_2}$-submodule and
$${\cal A}_k=\tilde{\cal H}_k\oplus(x_1^2+x_2x_5+x_3x_6+x_4x_7){\cal
A}_{k-2}.\eqno(1.15)$$ Moreover, an explicit basis for each
$\tilde{\cal H}_k$ was constructed in [X1]. Furthermore, Luo [L]
generalized the result (1.2) to certain noncanonical polynomial
representations of $o(n,\mbb{R})$.

To state our results in this paper, we denote
$$\vec x=\left(\begin{array}{c}x_1\\\vdots\\
x_n\end{array}\right),\qquad \vec f(\vec x)=\left(\begin{array}{c}f_1(\vec x)\\\vdots\\
f_n(\vec x)\end{array}\right),\eqno(1.16)$$
$$\hat{\cal
A}=\bigoplus_{k=0}^\infty\hat{\cal
A}_k\qquad\mbox{with}\;\;\hat{\cal A}_k=\{\vec f\mid f_j\in {\cal
A}_k\},\eqno(1.17)$$ where ${\cal A}_k$ is the space of polynomials
in $x_1,...,x_n$ of degree $k$. Moreover, we define
$$\hat{\cal H}_k=\{\vec f\in \hat{\cal A}_k\mid \iota_1\Delta
(\vec f)+(\iota_1+\iota_2)(\nabla^T\cdot\nabla )(\vec
f)=0\}.\eqno(1.18)$$ The action of the orthogonal group
$O(n,\mbb{R})$ on $\hat{\cal A}$ is defined by
$${\cal T}(\vec f(\vec x))=({\cal T}\vec f)({\cal T}^{-1}(\vec
x)).\eqno(1.19)$$ Furthermore, we denote
$\vs_r=(0,...,0,\stl{r}{1},0,...,0)^T$ and
$b=(\iota_1+\iota_2)/\iota_1$. Then $\vec f=\sum_{r=1}^nf_r\vs_r$.
We define  linear maps
$\psi,\vf_1,\vf_2:\mbb{R}[x_1,...,x_n]\rightarrow \hat{\cal A}$ by
$$\psi(x_{i_1}x_{i_2}\cdots x_{i_r})=\sum_{s=1}^rx_{i_1}\cdots
x_{i_{s-1}}x_{i_{s+1}}\cdots x_{i_r}\vs_s,\eqno(1.20)$$
$$\vf_1(x_{i_1}x_{i_2}\cdots x_{i_r})=\sum_{s=1}^rx_{i_1}\cdots
x_{i_{s-1}}x_{i_{s+1}}\cdots
x_{i_r}[r(\sum_{l=1}^nx_l^2)\vs_{i_s}-(2r+n-2)x_{i_s}\sum_{l=1}^nx_l\vs_l]\eqno(1.21)$$
and $\vf_2=(x_1^2+\cdots+x_n^2)\psi$.

Again we let ${\cal H}_k$ be the space of harmonic polynomials of
degree $k$. Set
$$\hat{\cal H}_{k,1}=\psi({\cal H}_{k+1}),\qquad \hat{\cal
H}_{k,2}=\{\sum_{l=1}^n f_l\vs_l\mid f_l\in{\cal
H}_k,\;\sum_{l=1}^nx_lf_l=0\},\eqno(1.22)$$
$$\hat{\cal
H}_{k,3}=\left(\vf_1+\frac{(2k+n-2)(k+n-3)(k-1)}{2(b^{-1}(2k+n-4)+k-1)}\vf_2\right)({\cal
H}_{k-1}).\eqno(1.23)$$ Denote
$d_{r,s}=x_{s}\partial_{x_{r}}-x_{r}\partial_{x_{s}}$ and set
$$\cal D=\left(\begin{array}{cccc}
                       0 & d_{3,4}&d_{4,2}&d_{2,3}\\
                       d_{4,3}& 0&d_{1,4}&d_{3,1}\\
                       d_{2,4}&d_{4,1}&0&d_{1,2}\\
                       d_{3,2}&d_{1,3}&d_{2,1}&0
                       \end{array}\right ).\eqno(1.24)$$
With $n=4,$ we define $$\hat{\mathcal H}_{k,2\pm}=\{\ \vec{f}\in
\hat{\mathcal {H}}_{k,2}\ \mid\ \mathcal{D}\vec{f}=\pm(k+1)\vec{f}\
\}.\eqno(1.25)$$

 \vspace{0.2cm}

\noindent{\bf Main Theorem 1}. {\it Let $n\geq 3$ be an integer. The
subspace of polynomial solutions $\hat{\cal H}_k=\hat{\cal
H}_{k,1}\oplus \hat{\cal H}_{k,2}\oplus \hat{\cal H}_{k,3}$ and
$$\hat{\cal A}_k=\hat{\cal H}_k\oplus (x_1^2+\cdots+x_n^2)\hat{\cal
A}_{k-2}.\eqno(1.24)$$Moreover, the subspaces $\hat{\cal
H}_{k,1},\;\hat{\cal H}_{k,2}\;(n\neq 4)$ and $\hat{\cal H}_{k,3}$
are irreducible $O(n,\mbb{R})$-submodules. When $n=4$, $\hat{\cal
H}_{k,2}=\hat{\cal H}_{k,2+}\oplus \hat{\cal H}_{k,2-}$ and
$\hat{\cal H}_{k,2+},\;\hat{\cal H}_{k,2-}$ are irreducible
$O(4,\mbb{R})$-submodules.}
 \vspace{0.2cm}

Furthermore, an explicit basis is constructed for each of the above
irreducible submodules.  Xu's Lemma in [X1] on the polynomial
solutions of constant-coefficient partial differential equation of
certain type is used to prove the equation $\hat{\cal H}_k=\hat{\cal
H}_{k,1}\oplus \hat{\cal H}_{k,2}\oplus \hat{\cal H}_{k,3}$. Our
above result is exactly a natural vector generalization of those
scalar ones in (1.2), (1.13) and (1.15). Indeed, it does not only
reveal the internal symmetry of Navier equations but also gives the
explicit realizations of certain irreducible $O(n,\mbb{R})$-modules.

Our second main result is obtaining explicit exact solutions of
Navier equations (1.6)
 subject to the initial conditions
$$\vec u(0,x_2,...,x_n)=\vec g_0(x_2,...,x_n),\;\;\;\vec u_{x_1}(0,x_2,...,x_n)=\vec g_1(x_2,...,x_n)
\eqno(1.25)$$ for $x_r\in[-a_r,a_r]$, and explicit exact solutions
of Lam\'{e} equations (1.7) subject to the initial conditions
$$\vec u(0,x_1,...,x_n)=\vec h_0(x_1,...,x_n),\;\;\;\vec u_t(0,x_1,...,x_n)=\vec h_1(x_1,...,x_n)
\eqno(1.26)$$ for $x_r\in[-b_r,b_r]$, where $\vec g_1,\vec g_2,\vec
h_1,\vec h_2$ are vector-valued continuous functions, and $a_r,b_r$
are positive real constants.  Xu's method of solving the
initial-value problems of flag partial differential equations in
[X1] and his matrix-differential-operator approach [X2] play
fundamental roles in obtaining our solutions. Moreover, Fourier
expansions are used.

Section 2 is devoted to the proof of Main Theorem 1. In Section 3,
we construct an explicit basis for each irreducible submodule
included in the solution space of Navier equations. Moreover, an
uniform explicit basis of the homogeneous polynomial solutions of
Navier equations is obtained by a method of Xu in [X1], whose
cardinality was pre-used in Section 2 to prove the completeness of
polynomial solutions for a technical reason. In Section 4, we solve
the above mentioned initial value problems.

\section{Polynomial Solutions and Representations}

In this section, we will study the homogeneous polynomial solutions
 of (1.6) in detail.

As we mentioned in the introduction, studying
$O(n,\mbb{R})$-representation structure of the polynomial solutions
is equivalent to studying their $o(n,\mbb{R})$-representation
structure via the exponential map from the Lie algebra to the Lie
group. Recall that $E_{r,s}$ is the square matrix with $1$ as its
$(r,s)$-entry and 0 as the others. The orthogonal Lie algebra
\begin{equation}o(n,\mbb{R})=\sum_{r,s=1}^n \mbb{R}
(E_{r,s}-E_{s,r}).\end{equation} Its action on the space $\hat{\cal
A}$ of polynomial vectors (cf. (1.17)) is given by
\begin{equation}(E_{r,s}-E_{s,r})(\vec f)=(x_r\ptl_{x_s}-x_s\ptl_{x_r})(\vec
f)+f_s\vs_r-f_r\vs_s, \end{equation} where
$\vs_r=(0,...,0,\stl{r}{1},0,...,0)^T$ as in the introduction.
Moreover, the elements of $o(n,\mbb{R})$ map solutions of (1.6) into
solutions. That is,
\begin{equation}\xi(\hat {\cal H}_k)\subset \hat {\cal
H}_k\qquad\for\;\;\xi\in o(n,\mbb{R}),\;k\in\mbb{N} \end{equation}
(cf. (1.18)). Note that $\vec{f}+\vec{g}i$ is a complex solution of
 Navier equations (1.6) if and only if $\vec{f}$ and $\vec{g}$ are real solutions.
In order to use the representation theory of the Lie algebras over
the complex field $\mbb{C}$,  we need to complexify our vector
spaces, indicated by the subindex of $\mbb{C}$. The complexification
of $o(n,\mbb{R})$ is $o(n,\mbb{C})$. We extend the representation
(2.2) of $o(n,\mbb{R})$ on $\hat{\cal A}$ to that of $o(n,\mbb{C})$
on $\hat{\cal A}_{\mbb{C}}$ $\mbb{C}$-bilinearly.

 Denote by $M_{m\times m}(\mbb{F})$ the algebra of $m\times m$
 matrices with entries in the field $\mbb{F}$.  Set
\begin{equation}o'(2m+1,\mbb{C})=\left\{\left(\begin{array}{ccc}
0&-\vec b^T&-\vec a^T\\
\vec a&A& A_1\\
\vec b& A_2&-A^T\end{array}\right)\mid A\in M_{m\times
m}(\mbb{C});A_1,A_2\in o(m,\mbb{C});\vec a,\vec
b\in\mbb{C}^m\right\}
\end{equation} and
\begin{equation}o'(2m,\mbb{C})=\left\{\left(\begin{array}{cc}A& A_1\\
A_2&-A^T\end{array}\right)\mid A\in M_{m\times
m}(\mbb{C});\;A_1,A_2\in o(m,\mbb{C})\right\}. \end{equation} Note
that $o'(3,\mbb{C})$ is the standard form of complex simple Lie
algebra of type $A_1$, $o'(4,\mbb{C})$ is the standard form of
complex semisimple  Lie algebra of type $A_1\oplus A_1$ and
$o'(6,\mbb{C})$ is the standard form of complex simple  Lie
algebra of type $A_3$ (or $D_3$). Moreover, $o'(2m+1,\mbb{C})$
with $m\geq 2$ is the standard form of the complex simple Lie
algebra of type $B_m$ and $o'(2m,\mbb{C})$ with $m\geq 4$ is the
standard form of the complex simple Lie algebra of type $D_m$. If
$n=2m+1$ is odd, we take
\begin{equation}H_{B_m}=\sum_{r=1}^m\mbb{C}(E_{r+1,r+1}-E_{m+r+1,m+r+1})\end{equation}
as a Cartan subalgebra of $o'(2m+1,\mbb{C})$,
\begin{eqnarray}& &\{E_{r+1,s+1}-E_{m+s+1,m+r+1},E_{r+1,m+s+1}-E_{s+1,m+r+1}\nonumber\\ & &E_{r_1,1}-E_{1,m+r_1}\mid
1\leq r<s\leq m;1\leq r_1\leq m\}\end{eqnarray} as positive root
vectors, and \begin{eqnarray}&
&\{E_{s+1,r+1}-E_{m+r+1,m+s+1},E_{m+s+1,r+1}-E_{m+r+1,s+1}\nonumber\\
& &E_{m+r_1,1}-E_{1,r_1}\mid 1\leq r<s\leq m;1\leq r_1\leq
m\}\end{eqnarray} as negative root vectors.

For $1\leq r\leq m$, we define the linear function $\varepsilon_r$
on $H^{*}_{B_{m}}$ by
\begin{eqnarray}
\varepsilon_{r}(E_{s+1,s+1}-E_{m+s+1,m+s+1})=\delta_{r,s}.
\end{eqnarray}
Then
$H^{*}_{B_{m}}=\mbox{span}\:\{\varepsilon_{1},\ldots,\varepsilon_{m}\}$
 and $\Phi=\{\pm\varepsilon_{r},\ \pm(\varepsilon_{r}\pm\varepsilon_{s})\mid r\neq
 s\}$ forms the root system of $o'(2m+1,\mbb{C})$. Denote by
 $\alpha_{r}=\varepsilon_{r}-\varepsilon_{r+1}$ for $r=1,\ldots,m-1$
 and $\alpha_{m}=\varepsilon_{m}$. As we know,
 $\{\alpha_{1},\ldots,\alpha_{m}\}$ is the simple root system.
Moreover, we define a symmetric bilinear form $(\cdot,\cdot)$ on
$H^{*}_{B_{m}}$ by
\begin{eqnarray}(\varepsilon_{r},\varepsilon_s)=\delta_{r,s}.\end{eqnarray}
Furthermore, we denote by $\lambda_{1},\ldots,\lambda_{m}$ the
 fundamental dominant weights in $H^{*}_{B_{m}}$, i.e.
 \begin{eqnarray}\la\lambda_{r},\alpha_{s}\ra=\frac{2(\lambda_{r},\alpha_{s})}{(\alpha_{s},\alpha_{s})}=\delta_{r,s}.
 \end{eqnarray}
 In particular,
 \begin{eqnarray}\varepsilon_{1}=\lambda_{1},\;
 \varepsilon_{2}=-\lambda_{1}+\lambda_{2},\;\ldots,\;
 \varepsilon_{m-1}=-\lambda_{m-2}+\lambda_{m-1},\;
\varepsilon_{m}=-\lambda_{m-1}+2\lambda_{m}.\end{eqnarray}

When $n=2m$ is even, we take
\begin{equation}H_{D_m}=\sum_{r=1}^m\mbb{C}(E_{r,r}-E_{m+r,m+r})\end{equation}
as a Cartan subalgebra of $o'(2m,\mathbb{C})$,
\begin{equation}\{E_{r,s}-E_{m+s,m+r},E_{r,m+s}-E_{s,m+r}\mid
1\leq r<s\leq m\}\end{equation} as positive root vectors, and
\begin{equation}\{E_{s,r}-E_{m+r,m+s},E_{m+s,r}-E_{m+r,s}\mid
1\leq r<s\leq m\}\end{equation} as negative root vectors. Moreover,
we define $\varepsilon_{r}\in H^{*}_{D_{m}}$ by
\begin{eqnarray}
\varepsilon_{r}(E_{s,s}-E_{m+s,m+s})=\delta_{r,s}.\end{eqnarray}
Similarly,
$H^{*}_{D_{m}}=\mbox{span}\:\{\varepsilon_{1},\ldots,\varepsilon_{m}\}$
 and $\Phi=\{\pm(\varepsilon_{r}\pm\varepsilon_{s})\mid r\neq
 s\}$ forms the root system of $o'(2m,\mbb{C})$. Moreover, we define a symmetric bilinear form $(\cdot,\cdot)$ on
$H^{*}_{D_{m}}$ by (2.10). Denote by
 $\alpha_{r}=\varepsilon_{r}-\varepsilon_{r+1}$ for $r=1,\ldots,m-1$
 and $\alpha_{m}=\varepsilon_{m-1}+\varepsilon_{m}$. Then
 $\{\alpha_{1},\ldots,\alpha_{m}\}$ is the simple root system.
 Moreover, $\lambda_{1},\ldots,\lambda_{m}$  are the
 fundamental dominant weights in $H^{*}_{D_{m}}$ defined by (2.11).
 In this case,
 \begin{eqnarray}\varepsilon_{1}=\lambda_{1},\;
 \varepsilon_{2}=-\lambda_{1}+\lambda_{2},\;\ldots,\;
 \varepsilon_{m-2}=-\lambda_{m-3}+\lambda_{m-2},\\
\varepsilon_{m-1}=-\lambda_{m-2}+\lambda_{m-1}+\lambda_{m},\;
\varepsilon_{m}=-\lambda_{m-1}+\lambda_{m}.\end{eqnarray}

Denote by $I_m$ the $m\times m$ identity matrix. Set
\begin{eqnarray}
K=\frac{1-i}{2}
 \left (\begin{array}{ccc}
       1+i & 0 & 0 \\
       0 & iI_{m} & I_{m}\\
       0 & I_{m} & iI_{m}
       \end{array} \right )
\end{eqnarray} if $n=2m+1$, and
\begin{eqnarray}K=\frac{1-i}{2}
 \left (\begin{array}{cc}
 iI_{m} & I_{m}\\
 I_{m} & iI_{m}
       \end{array} \right )
\end{eqnarray} when $n=2m$. Then $K$ is a symmetric matrix,
\begin{eqnarray}
K^{-1}=\frac{1+i}{2}
 \left (\begin{array}{ccc}
       1-i & 0 & 0 \\
       0 & -iI_{m} & I_{m}\\
       0 & I_{m} & -iI_{m}
       \end{array} \right )
\end{eqnarray} if $n=2m+1$, and
\begin{eqnarray}K^{-1}=\frac{1+i}{2}
 \left (\begin{array}{cc}
 -iI_{m} & I_{m}\\
 I_{m} & -iI_{m}
       \end{array} \right )
\end{eqnarray} when $n=2m$.
Furthermore, we have a Lie algebra isomorphism $\sgm:
o'(n,\mbb{C})\rightarrow o(n,\mbb{C})$ given by
\begin{equation}\sgm(X)=K^{-1}XK\qquad\for\;\;X\in o'(n,\mbb{C}).\end{equation}
We define a representation $\rho$ of $o'(n,\mbb{C})$ on $\hat{\cal
A}$ by
\begin{equation}\rho(X) (\vec f)=\sgm(X)(\vec f)\qquad\for\;\;X\in
o'(n,\mbb{C}).\end{equation} So $\hat{\cal A}$ forms an
$o'(n,\mbb{C})$-module.

In order to study the $o'(n,\mbb{C})$-module structure of ${\cal
A}_{\mbb{C}}$, we let
\begin{eqnarray}
\left (\begin{array}{c}
y_{1} \\
\vdots \\
y_{n}
\end{array} \right )=
\frac{1+i}{2}K^{-1} \left (\begin{array}{c}
x_{1} \\
\vdots \\
x_{n}
\end{array} \right ).
\end{eqnarray}
Define two representations $\rho_1$ and $\rho_2$ of $gl(n,\mbb{C})$
on ${\cal A}_{\mbb{C}}=\mbb{C}[x_1,...,x_n]=\mbb{C}[y_1,...,y_n]$ by
\begin{equation}\rho_1(E_{r,s})=y_r\ptl_{y_s},\qquad
\rho_2(E_{r,s})=x_r\ptl_{x_s}. \end{equation} Writing
$K=(b_{r,s})_{n\times n}$ and $K^{-1}=(c_{r,s})_{n\times n}$, we
have
\begin{equation}
K^{-1}E_{p,q}K=\sum_{r,s=1}^n c_{s,p}b_{q,r}E_{s,r}
\end{equation}
 and
\begin{equation}
y_{p}\partial_{y_{q}}=
 \sum_{r, s=1}^n
   \frac{\partial y_{p}}{\partial x_{s}}
   \frac{\partial x_{r}}{\partial y_{q}}
 x_{s}\partial_{x_{r}}=
\sum_{r,s=1}^n c_{s,p}b_{q,r}x_{s}\partial_{x_{r}}.
\end{equation}
Thus
\begin{equation}\rho_1(X)=\rho_2(K^{-1}XK)\qquad\for\;\;X\in
gl(n,\mbb{C}).\end{equation} Recall
$\vs_r=(0,...,0,\stl{r}{1},0,...,0)^T$. Furthermore, we set
\begin{eqnarray}
\left (\begin{array}{c}
\kappa_{1} \\
\vdots \\
\kappa_{n}
\end{array} \right )=
\frac{1+i}{2}K^{-1} \left (\begin{array}{c}
\vs_{1} \\
\vdots \\
\vs_n
\end{array} \right ).
\end{eqnarray}
Then
\begin{equation}\hat{\cal A}_{\mbb{C}}=\sum_{r=1}^n{\cal
A}_{\mbb{C}}\:\vs_r=\sum_{s=1}^n{\cal
A}_{\mbb{C}}\:\kappa_s.\end{equation}

We define  two representations $\hat\rho_1$ and $\hat\rho_2$ of
$gl(n,\mbb{C})$ on $\hat{\cal A}_{\mbb{C}}$ by
\begin{equation}\hat\rho_1(X)(\sum_{r=1}^nf_r\kappa_r)=\sum_{r=1}^n\rho_1(X)(f_r)\kappa_r
+(f_1,...,f_n)X^T\left(\begin{array}{c}\kappa_1\\\vdots\\
\kappa_n\end{array}\right),\end{equation}
\begin{equation}\hat\rho_2(X)(\sum_{r=1}^ng_r\vs_r)=\sum_{r=1}^n\rho_2(X)(g_r)\vs_r
+(g_1,...,g_n)X^T\left(\begin{array}{c}\vs_1\\\vdots\\
\vs_n\end{array}\right),\end{equation} for $f_r,g_r\in {\cal
A}_{\mbb{C}}$ and $X\in gl(n,\mbb{C})$. By (2.27)-(2.31), we have
\begin{equation}\hat\rho_1(X)=\hat\rho_2(K^{-1}XK)\qquad\for\;\;X\in
gl(n,\mbb{C}).\end{equation} According to (2.2), (2.23), (2.24) and
(2.34), we obtain:

\begin{lemma} For $X\in o'(n,\mbb{C})$, $\rho(X)=\hat\rho_1(X)$. \end{lemma}

Denote by ${\cal A}_{\mbb{C},k}$ the subspace of the polynomials
with degree $k$ in ${\cal A}_{\mbb{C}}$. Set
\begin{equation} V=\mbb{R}^n=\bigoplus_{s=1}^n\mbb{R}\vs_s,\qquad
V_{\mbb{C}}=\mbb{C}^n=\bigoplus_{r=1}^n\mbb{C}\kappa_r=\bigoplus_{s=1}^n\mbb{C}\vs_s,\end{equation}
\begin{equation}{\cal H}'_k=\{f\in {\cal A}_{\mbb{C},k}\mid
(\ptl_{y_1}^2+2\sum_{r=1}^m\ptl_{y_{r+1}}\ptl_{y_{m+r+1}})(f)=0\}\end{equation}
if $n=2m+1$, and
\begin{equation}{\cal H}'_k=\{f\in {\cal A}_{\mbb{C},k}\mid
(\sum_{r=1}^m\ptl_{y_r}\ptl_{y_{m+r}})(f)=0\}\end{equation} if
$n=2m$. It can be verified that
\begin{equation} {\cal H}'_k=\mbb{C}\otimes_{\mbb{R}}{\cal
H}_k=\{f\in {\cal A}_{\mbb{C},k}\mid
(\sum_{r=1}^n\ptl_{x_r}^2)(f)=0\}\end{equation} (also see (2.68)).
Now $V$ forms an $o(n,\mbb{R})$-submodule of $\hat{\cal A}$ that
gives the basic representation with the canonical basis
$\{\vs_1,...,\vs_n\}$ (cf. (2.2)). Moreover, $V_{\mbb C}$ forms an
$o'(n,\mbb{C})$-submodule of $\hat{\cal A}_{\mbb{C}}$ that gives the
basic representation with the canonical basis
$\{\kappa_1,...,\kappa_n\}$ (cf. (2.30)). Furthermore, ${\cal A}$
forms an $o(n,\mbb{R})$-module with the representation
$\rho_2|_{o(n,\mbb{R})}$, and ${\cal A}_{\mbb{C}}$ forms an
$o'(n,\mbb{C})$-module with the representation
$\rho_1|_{o'(n,\mbb{C})}$ (cf. (2.26)). As a real
$o(n,\mbb{R})$-module, $\hat{\cal
A}_k={\cal{A}}_{k}V\cong{\cal{A}}_{k}\otimes_{\mbb{R}} V$ (cf.
(2.2)). On the other hand, $\hat{\cal
A}_{\mbb{C},k}={\cal{A}}_{\mbb{C}, k}V_{\mbb{C}}\cong
{\cal{A}}_{\mbb{C}, k}\otimes_{\mbb{C}} V_{\mbb{C}}$ as a complex
$o'(n,\mbb{C})$-module by the above lemma. We want to decompose
$\hat{\cal A}_k$ as a direct sum of real irreducible
$o(n,\mbb{R})$-submodules via decomposing $\hat{\cal A}_{\mbb C,k}$
as a direct sum of complex irreducible $o'(n,\mbb{C})$-submodules.
To study the complex $o'(n,\mbb{C})$-module ${\cal{A}}_{\mbb{C},
k}\bigotimes_{\mbb{C}} V_{\mbb{C}}$, we recall a fact about the
tensor product of modules for a finite-dimensional complex
semisimple Lie algebra ${\cal G}$.

Denote by $V(\lambda)$ a finite-dimensional  irreducible ${\cal
G}$-module with highest weight $\lambda$, by $\Lambda$ the weight
lattice of ${\cal G}$ and by $\Lambda^{+}$ the set of dominant
weights. Moreover,  $V_{\nu}(\lambda)$ stands for the weight space
of $V(\lambda)$ with the weight $\nu$.

\begin{lemma} (e.g., cf. [GE])
The irreducible representations occurring in
 $V(\lambda)\bigotimes V(\mu)$
 have highest weights of the form
$\mu+\nu\in\Lambda^{+}$, where $\nu$ is a weight in $V(\lambda)$.
Moreover, the multiplicity of $V(\mu+\nu)$ in $V(\lambda)\bigotimes
V(\mu)$ is less than or equal to ${\it dim}\: V_{\nu}(\lambda)$.
\end{lemma}

 The weights of $o'(2m,\mbb{C})$-module $V_{\mbb{C}}\cong V(\lambda_1)$ with $m\geq 3$  are
\begin{eqnarray}
\varepsilon_{1}\succ \varepsilon_{2}\succ \cdots
\varepsilon_{m-2}\succ\varepsilon_{m-1}\begin{array}{l}\succ
\varepsilon_m\\\succ-\varepsilon_m\end{array}\succ-\varepsilon_{m-1}
\succ\cdots\succ-\varepsilon_{1}.
\end{eqnarray}
Moreover, the weights of $o'(2m+1,\mbb{C})$-module $V_{\mbb{C}}\cong
V(\lambda_1)$ with $m\geq 2$  are
\begin{eqnarray}
\varepsilon_{1}\succ \varepsilon_{2}\succ \ldots
\varepsilon_{m-2}\succ\varepsilon_{m-1}\succ\varepsilon_{m}\succ0
\succ-\varepsilon_{m}\succ-\varepsilon_{m-1}\succ\ldots\succ-\varepsilon_{2}\succ-\varepsilon_{1}.
\end{eqnarray}
The weights of $o'(4,\mbb{C})$-module $V_{\mbb{C}}\cong
V(\lambda_1+\lambda_2)$ are
\begin{eqnarray}
\lambda_1+\lambda_2\begin{array}{l}\succ
-\lambda_1+\lambda_2\\\succ\lambda_1-\lambda_2\end{array}\succ-\lambda_1-\lambda_2.
\end{eqnarray}
The weights of $o'(3,\mbb{C})$-module $V_{\mbb{C}}\cong
V(2\lambda_1)$ are
\begin{eqnarray}
2\lambda_1\succ0\succ-2\lambda_1.
\end{eqnarray}
Each weight in $V_\mathbb{C}$ is with multiplicity 1 (cf. (2.9) and
(2.16)).

It is well known that
\begin{eqnarray}{\cal{A}}_{\mbb{C},
k}={\cal H}'_k\oplus
(y_1^2+2\sum_{r=1}^{2m}y_{r+1}y_{m+r+1}){\cal{A}}_{\mbb{C},
k-2}\end{eqnarray} if $n=2m+1$, and
\begin{eqnarray}{\cal{A}}_{\mbb{C},
k}={\cal H}'_k\oplus (\sum_{r=1}^{2m}y_ry_{m+r}){\cal{A}}_{\mbb{C},
k-2}\end{eqnarray} if $n=2m$. Moreover,
$y_1^2+2\sum_{r=1}^{2m}y_{r+1}y_{m+r+1}$ is an
$o'(2m+1,\mbb{C})$-invariant and $\sum_{r=1}^{2m}y_ry_{m+r}$ is an
$o'(2m,\mbb{C})$-invariant. Furthermore, ${\cal H}'_k\cong
V(k\lambda_1)$ if $n>4 $, ${\cal H}'_k\cong V(2k\lambda_1)$ if $n=3$
and ${\cal H}'_k\cong V(k\lambda_1+k\lambda_2)$ if $n=4$.

\begin{lemma}
i) As $o'(n,\mbb{C})$-modules with $n\geq 7$,
\begin{equation}
{\cal H}_k'V_{\mbb{C}}\cong{\cal H}_k'\otimes_{\mbb{C}}
V_{\mbb{C}}\cong V((k+1)\lambda_{1})\oplus
V((k-1)\lambda_{1}+\lambda_{2})\oplus V((k-1)\lambda_{1}).
\end{equation}
ii) For $o'(6,\mbb{C})$,
\begin{equation}
{\cal H}_k'V_{\mbb{C}}\cong{\cal H}_k'\otimes_{\mbb{C}}
V_{\mbb{C}}\cong V((k+1)\lambda_{1})\oplus
V((k-1)\lambda_{1}+\lambda_{2}+\lambda_{3})\oplus
V((k-1)\lambda_{1}).
\end{equation}
iii) For $o'(5,\mbb{C})$,
\begin{equation}
{\cal H}_k'V_{\mbb{C}}\cong{\cal H}_k'\otimes_{\mbb{C}}
V_{\mbb{C}}\cong V((k+1)\lambda_{1})\oplus
V((k-1)\lambda_{1}+2\lambda_{2})\oplus V((k-1)\lambda_{1}).
\end{equation}
iv) For $o'(4,\mbb{C})$,
\begin{eqnarray}
&&{\cal H}_k'V_{\mbb{C}}\cong{\cal H}_k'\otimes_{\mbb{C}}
V_{\mbb{C}} \nonumber\\&\cong&
V((k+1)(\lambda_{1}+\lambda_{2}))\oplus
V((k-1)\lambda_{1}+(k+1)\lambda_{2})\oplus
V((k+1)\lambda_{1}+(k-1)\lambda_{2})\nonumber\\&&\oplus
V((k-1)(\lambda_{1}+\lambda_{2})).
\end{eqnarray}
v) For $o'(3,\mbb{C})$,
\begin{equation}
{\cal H}_k'V_{\mbb{C}}\cong{\cal H}_k'\otimes_{\mbb{C}}
V_{\mbb{C}}\cong V(2(k+1)\lambda_{1})\oplus V(2k\lambda_{1})\oplus
V(2(k-1)\lambda_{1}).
\end{equation}
\end{lemma}
{\it Proof}. i)  First we consider the case  that $n=2m$ is even
with $m\geq4$. Lemma 2.2, (2.17), (2.18) and (2.39) tell us that the
irreducible modules may occur in $V(k\lambda_{1})\otimes
V(\lambda_{1})$ are $ V((k+1)\lambda_{1})$,
$V((k-1)\lambda_{1}+\lambda_{2})$ and $V((k-1)\lambda_{1})$. The
multiplicities of them are all less than or equal to 1. According to
the dimension formula of finite-dimensional irreducible modules for
a complex finite-dimensional simple Lie algebra (e.g., cf. Page 139
in [H]), we have
\begin{eqnarray}
&&\dim V((k-1)\lambda_{1}+\lambda_{2})\nonumber\\&=&
\frac{\prod\limits_{\alpha\succ0}((k-1)\lambda_{1}+\lambda_{2}+\delta,\
\alpha)} {\prod\limits_{\alpha\succ0}(\delta,\
\alpha)}\nonumber\\&=&
\frac{k(k+2m-2)\prod\limits_{s=3}^{m}(k+2m-1-s)(s-1)
\prod\limits_{j=3}^{m}(2m-1-j)(j-1)}
{(2m-3)\prod\limits_{s=3}^{m}(2m-1-s)(s-1)
\prod\limits_{j=3}^{m}(2m-2-j)(j-2)}\nonumber\\&=&
\frac{2k(k+2m-2)(k+m-1)(k+2m-4)!}{(2m-3)!(k+1)!}\nonumber\\
&=&\frac{k(k+n-2)(2k+n-2)(k+n-4)!}{(n-3)!(k+1)!},
\end{eqnarray}
where $\delta=\sum_{r=1}^m\lambda_r$. Moreover,
\begin{eqnarray}
&&\dim(V(k\lambda_{1})\otimes
V(\lambda_{1}))-\dim V((k+1)\lambda_{1})-\dim V((k-1)\lambda_{1})\nonumber\\
&=&n{k+n-3\choose n-2}+n{k+n-2\choose n-2}-{k+1+n-3\choose
n-2}\nonumber\\&&-
{k+1+n-2\choose n-2}-{k-1+n-3\choose n-2}-{k-1+n-2\choose n-2}\nonumber\\
&=&\frac{k(k+n-2)(2k+n-2)(k+n-4)!}{(n-3)!(k+1)!}\nonumber\\
&=&\dim V((k-1)\lambda_{1}+\lambda_{2}).
\end{eqnarray}
Hence all the three modules occur in ${\cal H}_k'\otimes_{\mbb{C}}
V_{\mbb{C}}$.

Next we consider the case that $n=2m+1$ is odd with $m\geq 3$. The
irreducible modules that may occur in $V(k\lambda_{1})\otimes
V(\lambda_{1})$ are $ V((k+1)\lambda_{1})$, $V(k\lambda_{1})$,
$V((k-1)\lambda_{1}+\lambda_{2})$ and $V((k-1)\lambda_{1})$. Their
multiplicities are all less than or equal to 1 by Lemma 2.2, (2.12)
and (2.40). Note that $ V((k+1)\lambda_{1})$ must occur. Since
\begin{eqnarray}
&&\dim V((k-1)\lambda_{1}+\lambda_{2})\nonumber\\&=&
\frac{\prod\limits_{\alpha\succ0}((k-1)\lambda_{1}+\lambda_{2}+\delta,\
\alpha)} {\prod\limits_{\alpha\succ0}(\delta,\ \alpha)}\nonumber\\
&=&
\frac{k(k+2m-1)(k+m-\frac{1}{2})\prod\limits_{s=3}^{m}(k+2m-s)(k+s-1)
\prod\limits_{j=3}^{m}(2m-j)(j-1)}
{(2m-2)(m-\frac{3}{2})\prod\limits_{s=3}^{m}(2m-s)(s-1)
\prod\limits_{j=3}^{m}(2m-1-j)(j-2)} \nonumber\\&=&
\frac{2k(k+2m-1)(k+m-\frac{1}{2})(k+2m-3)!}{(2m-2)!(k+1)!}\nonumber\\
&=&\frac{k(k+n-2)(2k+n-2)(k+n-4)!}{(n-3)!(k+1)!}\nonumber\\
&=&\dim(V(k\lambda_{1})\otimes V(\lambda_{1}))-\dim
V((k+1)\lambda_{1})-\dim
V((k-1)\lambda_{1}),\;\;\;\;\;\;\;\;\;\;\;\;\;\;
\end{eqnarray}
at most two of $\{V(k\lambda_{1}),\
V((k-1)\lambda_{1}+\lambda_{2}),\ V((k-1)\lambda_{1})\}$ occur. So
(2.45) with $n=2m+1$ follows from the fact
\begin{eqnarray}
\dim V((k-1)\lambda_{1}+\lambda_{2})>\dim V(k\lambda_{1})>\dim
V((k-1)\lambda_{1}).
\end{eqnarray}

 ii) It follows by a similar argument as that of $n=2m$ in i).

iii) It is obtained by a similar argument as that of $n=2m+1$ in i).

iv) The irreducible modules may occur in ${\cal H}'_k\otimes
V_{\mbb{C}}$ are the four ones in the right side of (2.48) by lemma
2.2. The conclusion follows from the fact
 \begin{eqnarray}
 \dim
V((k+1)(\lambda_{1}+\lambda_{2}))=(k+2)^{2},\;\dim
V((k-1)(\lambda_{1}+\lambda_{2}))=k^{2},
\end{eqnarray}
\begin{eqnarray}
&&\dim V((k+1)(\lambda_{1})+(k-1)\lambda_{2}))\nonumber\\&=& \dim
V((k-1)(\lambda_{1})+(k+1)\lambda_{2}))=k(k+2)
\end{eqnarray} and
\begin{eqnarray}
\dim V(k(\lambda_{1}+\lambda_{2}))\otimes
V(\lambda_{1}+\lambda_{2})=4(k+1)^{2}.
\end{eqnarray}

v) The set of weights occurring in $V(2\lambda_1)$ of
$o'(3,\mbb{C})$ is $\Pi(\lambda_1)=\{2\lambda_1,0,-2\lambda_1\}$.
The statement holds because  $\dim V(2k\lambda_1)=2k+1$ for any
$k\geq1$.   $\qquad\Box$ \vspace{0.4cm}

For convenience, we treat $\kappa_r$ and $\varsigma_r$ as variables.
With these notations, we may write
\begin{equation}
(E_{r,s}-E_{s,r})(\sum_{j=1}^{n}f_{j}\varsigma_{j})=
(x_{r}\partial_{x_{s}}-x_{s}\partial_{x_{r}}+
\varsigma_{r}\partial_{\varsigma_{s}}-\varsigma_{s}\partial_{\varsigma_{r}})
(\sum_{j=1}^{n}f_{j}\varsigma_{j})
\end{equation} for $E_{r,s}-E_{s,r}\ \in o(n,\mathbb{R}).$

 We define the bar operation on
$\hat{\mathcal{A}}_\mathbb{C}$ by
\begin{eqnarray}
\overline{\sum\limits_{r_1,\ldots,r_n}^{\infty}\sum\limits_{j=1}^{n}
a_{r_1,\ldots,r_n,j}x_1^{1}\cdots x_n^{n}\varsigma_j}=
\sum\limits_{r_1,\ldots,r_n}^{\infty}\sum\limits_{j=1}^{n}\overline{a_{r_1,\ldots,r_n,j}}
x_1^{1}\cdots
x_n^{n}\varsigma_j,\;\;\;\;a_{r_1,\ldots,r_n,j}\in\mathbb{C}.
\end{eqnarray}

For a transformation $T$ on $\hat{\mathcal{A}}_\mathbb{C}$, we
define its conjugate operator by
\begin{eqnarray}
\overline{T}(\vec f)=\overline{T(\vec f)}\;\;\;\;\;\;\;\;\mbox{for}\
\vec f\in\hat{\mathcal{A}}_\mathbb{C}.
\end{eqnarray}

For expository convenience, we shift the sub-indices by $-1$ when
$n=2m+1$ is odd; for instance,
\begin{eqnarray}
x_r\rightarrow x_{r-1},\;\; y_r\rightarrow y_{r-1},\;\;
\varsigma_r\rightarrow \varsigma_{r-1},\;\; \kappa_r\rightarrow
\kappa_{r-1},\;\;E_{r,s}\rightarrow E_{r-1,s-1}
\end{eqnarray}
for $1\leq r,s\ \leq 2m+1$.

In this way, we always have
\begin{eqnarray}
y_r=\frac{x_r+ix_{m+r}}{2},\;y_{m+r}=\frac{ix_r+x_{m+r}}{2}
\end{eqnarray} and
\begin{eqnarray}
\kappa_r=\frac{\varsigma_r+i\varsigma_{m+r}}{2},\;
\kappa_{m+r}=\frac{i\varsigma_r+\varsigma_{m+r}}{2}
\end{eqnarray} for $1\leq r\leq m$ by (2.25) and (2.30) in the both cases of
$n=2m$ and $n=2m+1$. Moreover,
\begin{eqnarray}
y_0=\frac{1+i}{2}x_0\;\;\mbox{and}\;\;
\kappa_0=\frac{1+i}{2}\varsigma_0\;\;\;\;\mbox{if}\ n=2m+1.
\end{eqnarray}

Note that
\begin{eqnarray}
y_{m+r}=i\overline{y_r}\;\; \mbox{and}\;\;
\kappa_{m+r}=i\overline{\kappa_r}\;\; \for \;\;1\leq r\leq m,
\end{eqnarray}
also
\begin{eqnarray}
\overline{y_0}=-iy_0\;\;\mbox{and}\;\;\overline{\kappa_0}=-i\kappa_0\;\;
\mbox{if}\;\;n=2m+1.
\end{eqnarray}

Furthermore, (2.61) yields
\begin{eqnarray}
\partial_{x_r}=\frac{\partial_{y_r}+i\partial_{y_{m+r}}}{2},\;\;
\partial_{x_{m+r}}=\frac{i\partial_{y_r}+\partial_{y_{m+r}}}{2}
\end{eqnarray} and (2.63) says
\begin{equation}
\partial_{x_0}=\frac{1+i}{2}\partial_{y_0}.
\end{equation}

Thus the Laplacian operator
\begin{eqnarray}
\Delta=\sum\limits_{r=1}^{n}\partial_{x_r}^{2}=\left\{\begin{array}{l}
i\sum\limits_{r=1}^{m}\partial_{y_r}\partial_{y_{m+r}}\;\;\mbox{if}\;\;n=2m;\\
i(\frac{1}{2}\partial_{y_0}^{2}+\sum\limits_{r=1}^{m}\partial_{y_r}\partial_{y_{m+r}})
\;\;\mbox{if}\;\;n=2m+1.\end{array}\right.
\end{eqnarray}

In particular, (2.38) holds. Expressions (2.64)-(2.67) also imply
\begin{eqnarray}
\overline{\partial_{y_s}}=\partial_{\overline{y_s}},\ \ \ \ \ \ \ \
\overline{\partial_{\kappa_s}}=\partial_{\overline{\kappa_s}}\ \ \ \
\mbox{for}\ 1\leq s\leq n.
\end{eqnarray}

A complex module $W$ of $o(n,\mathbb{R})$ is of {\em real type} if
there exists a real module $W_0$ of $o(n,\mathbb{R})$ satisfying
$W=W_0\bigotimes_{\mathbb{R}}\mathbb{C}$. In this case, we call
$W_0$ the {\em real form} of $W$.

 Because we
have to deal with the real module finally, we give a lemma to
describe how to find the real form of a module in ${\mathcal H}_k'
V_{\mbb{C}}$. Recall that we have chosen a Cartan subalgebra and
positive (negative) root vectors for $o'(n,\mbb{C})$ in (2.6)-(2.9)
and (2.14)-(2.16) with sub-indices shifted by $-1$. A \emph{highest\
(lowest) \ weight\ vector} of an $o'(n,\mbb{C})$-module is a weight
vector nullified by positive (negative) root vectors.

\begin{lemma} Suppose that  $W\subset{\mathcal H}_k'
V_{\mbb{C}}$ is an irreducible submodule of $o'(n,\mathbb{C})$. Then
$W_0=\{\vec{h}\in W\mid \overline{\vec h}=\vec h\}$, which is an
irreducible real $o(n,\mathbb{R})$-module, is the real form of $W$.
\end{lemma}
{\it Proof}. Recall the action of $o'(n,\mathbb{C})$ on
$\hat{\mathcal{A}}_{\mbb{C}}$ by $\hat{\rho}_1$ given in (2.32).
Observe that
\begin{eqnarray}
&&\overline{y_{s}\partial_{y_{r}}-y_{m+r}\partial_{
y_{m+s}}+\kappa_{s}\partial_{\kappa_{r}}-\kappa_{m+r}\partial_{\kappa_{m+s}}}\nonumber\\&=&
-(y_{r}\partial_{y_{s}}-y_{m+s}\partial_{
y_{m+r}}+\kappa_{r}\partial_{\kappa_{s}}-\kappa_{m+s}\partial_{\kappa_{m+r}}),
\end{eqnarray}
\begin{eqnarray}
&&\overline{y_{m+s}\partial_{y_{r}}-y_{m+r}\partial_{
y_{s}}+\kappa_{m+s}\partial_{\kappa_{r}}-\kappa_{m+r}\partial_{\kappa_{s}}}\nonumber\\&=&
-(y_{r}\partial_{y_{m+s}}-y_{s}\partial_{
y_{m+r}}+\kappa_{r}\partial_{\kappa_{m+s}}-\kappa_{s}\partial_{\kappa_{m+r}}).
\end{eqnarray}

Moreover,
\begin{eqnarray}
&&\overline{y_{r}\partial_{y_{0}}-y_{0}\partial_{
y_{m+r}}+\kappa_{r}\partial_{\kappa_{0}}-\kappa_{0}\partial_{\kappa_{m+r}}}\nonumber\\&=&
-(y_{0}\partial_{y_{r}}-y_{m+r}\partial_{
y_{0}}+\kappa_{0}\partial_{\kappa_{r}}-\kappa_{m+r}\partial_{\kappa_{0}})
\end{eqnarray} if $n=2m+1.$ Note for a positive root vector $A\in
o'(n,\mathbb{C})$, its transpose $A^T$ is a negative root vector.
The above expressions says that
\begin{eqnarray}
A^T(\overline{\vec{f}})=-\overline{A(\vec{f})}.
\end{eqnarray}
for $\vec{f}\in\hat{\mathcal{A}}_{\mathbb{C}}$.

Now let $\vec{f}$ be a highest weight vector of $W$ with weight
$\lambda$. Then $\overline{\vec{f}}$ is a lowest weight vector of
some submodule of ${\mathcal H}_k'V_{\mbb{C}}$ by (2.70)-(2.72). By
(2.36), (2.37) and (2.68), $\overline{{\mathcal H}_k'}={\mathcal
H}_k'$. Moreover, (2.64) and (2.65) imply the weight subspaces
\begin{equation}\overline {({\mathcal H}'_k)_{\mu_1}}=({\mathcal
H}_k')_{-\mu_1},\qquad
\overline{(V_{\mbb{C}})_{\mu_2}}=(V_{\mbb{C}})_{-\mu_2}.\end{equation}
Thus
\begin{equation}({\mathcal H}_k'
V_{\mbb{C}})_{\mu}=\bigoplus_{\mu_1+\mu_2=\mu} ({\mathcal
H}'_k)_{\mu_1} (V_{\mbb{C}})_{\mu_2}\Rightarrow \overline{({\mathcal
H}_k' V_{\mbb{C}})_{\mu}}=({\mathcal
H}'_kV_{\mbb{C}})_{-\mu}.\end{equation} So $\overline{\vec{f}}$ is
of weight $-\lambda$. Since the lowest weight of $W$ is also
$-\lambda$ by the highest weights listed in Lemma 2.3, the
$o'(n,\mathbb{C})$-submodule generated by $\overline{\vec{f}}$ must
be isomorphic to $W$. But the multiplicity of any
$o'(n,\mathbb{C})$-irreducible submodule in ${\mathcal H}_k'
V_{\mbb{C}}$ is 1 by Lemma 2.3. Therefore, $\overline{\vec{f}}\in
W$.

Recall that $W$ is spanned by the elements of the form:
\begin{eqnarray}
\vec{g}=f_{r_1}^{a_1}f_{r_2}^{a_2}\cdots f_{r_m}^{a_m}\vec{f}
\end{eqnarray}
where $f_{r_j}$ is the operator of the $r_j$th negative simple root
vector in $o'(n,\mbb{C}),$ $a_r\in \mathbb{N}$ and $1\leq r_j\leq m$
for $n=2m$ or $n=2m+1$. Then
\begin{eqnarray}
(-1)^{\sum\limits_{r=1}^{m}a_r}\overline{\vec{g}}&=&(-1)^{\sum\limits_{r=1}^{m}a_r}\overline{f_{r_1}^{a_1}f_{r_2}^{a_2}\cdots
f_{r_m}^{a_m}\vec{f}}\nonumber\\&=& (-1)^{\sum\limits_{r=1}^{m}a_r}
\overline{f_{r_1}^{a_1}}\
\overline{f_{r_2}^{a_2}}\cdots\overline{f_{r_m}^{a_m}}
\overline{\vec{f}}\nonumber\\&=& e_{r_1}^{a_1}e_{r_2}^{a_2}\cdots
e_{r_m}^{a_m}\overline{\vec{f}},
\end{eqnarray}
where $e_{r_j}$ is the operator of the $r_j$th positive simple root
vector in $o'(n,\mbb{C})$.  It follows that the real and imaginary
parts of $\vec{g}$ are in $W$ whenever $\vec{g}\in W$. This shows
$W=W_0+iW_0$. By Lemma 2.1, $W_0$ must be a real irreducible
$o(n,\mathbb{R})$-submodule. $\qquad\Box$ \vspace{0.4cm}

Now we want to find the highest weight vectors of the direct
summands of
 ${\mathcal {H}}'_kV_{\mbb{C}}$ in Lemma 2.3.
 We denote
 ${\mathcal {H}}'_kV_{\mbb{C}}=W_{1}\oplus W_{2}\oplus
 W_{3}^{'}$,
 where $W_{j}$ (or $W_{j}^{'}$) is correspondence to the $j$th direct summand in
 every equalities in Lemma 2.3 expect the case $n=4$. Let $W_{2}=W_{2,-}\bigoplus W_{2,+}$ be the
 sum of the middle two submodules if $n=4$. According to Lemma 2.4,
 we have real irreducible $o(n,\mathbb{R})$-submodules
 $\hat{\cal H}_{k,1}, \hat{\cal H}_{k,2}, \hat{\cal H}_{k,3}^{'}$
 such that $W_{j}=\hat{\cal H}_{k,j}+i\hat{\cal H}_{k,j}$ for $j=1,2$ and
 $W_{3}^{'}=\hat{\cal H}_{k,3}^{'}+i\hat{\cal H}_{k,3}^{'}$. In
 particular,  $\mathcal {H}_{k}V=
 \hat{\cal H}_{k,1}\oplus \hat{\cal H}_{k,2}\oplus \hat{\cal
 H}_{k,3}^{'}$ by (2.35) and (2.38).  in the case of $n=4$,
 $\hat{\cal H}_{k,2\mp}$ are the similar real forms of $W_{2,\mp}$.

\begin{lemma} i) The following vectors $v_{j}\ (res.\ v'_{3})$  of $W_{j}\ (res.\
W'_{3})$ and $v_{2,\mp}$ of $W_{2,\mp}$ are highest weight vectors:
\begin{equation}
v_{1}=y_{1}^{k}\kappa_{1}=\frac{1}{2^{k+1}}(x_{1}+ix_{m+1})^{k}(\varsigma_{1}+i\varsigma_{m+1}),
\end{equation}
\begin{equation}
v_{2}=-y_{2}y_{1}^{k-1}\kappa_{1}+y_{1}^{k}\kappa_{2}\ \ \ \ \ \ \
 \ \ for\ n>4,
\end{equation}
\begin{equation}
v_{2,-}=-y_{2}y_{1}^{k-1}\kappa_{1}+y_{1}^{k}\kappa_{2},\ \
v_{2,+}=-y_{4}y_{1}^{k-1}\kappa_{1}+y_{1}^{k}\kappa_{4}\ \ \ for\
n=4,
\end{equation}
\begin{equation}
v_{2}=-y_{0}y_{1}^{k-1}\kappa_{1}+y_{1}^{k}\kappa_{0}\ \ \ \ \ \ \ \
\ \ \ \ for\ n=3,
\end{equation}
\begin{eqnarray}
v_{3}^{'}=2(k-1)\sum_{j=1}^{m}y_{j}y_{m+j}y_{1}^{k-2}\kappa_{1}
-(2k+n-4)\sum_{j=1}^{m}y_{1}^{k-1}(y_{m+j}\kappa_{j}+y_{j}\kappa_{m+j})
\end{eqnarray} for $n=2m,$
\begin{eqnarray}
\nonumber v_{3}^{'} &=&-(2k+n-4)y_{0}y_{1}^{k-1}\kappa_{0}
+2(k-1)(\sum_{j=1}^{m}y_{j}y_{m+j}+\frac{1}{2}y_{0}^{2})y_{1}^{k-2}\kappa_{1}\\
&&-(2k+n-4)\sum_{j=1}^{m}y_{1}^{k-1}(y_{m+j}\kappa_{j}+y_{j}\kappa_{m+j})
\end{eqnarray} for $n=2m+1.$

ii)The vectors $v_{1}$, $v_{2}$ and $v_{2,\mp}$ are solutions of
Navier equations (1.6). Moreover,  $\hat{\cal H}_{k,1}$ and
$\hat{\cal H}_{k,2}$ are subspaces of the solution space.
\end{lemma}
{\it Proof}. (a) We want to find the highest weight vectors of
$W_{1}$. The action of positive root vectors of $o'(2m,\mbb{C})$ on
$\hat{\mathcal {A}}_{\mbb{C}}$ are operators:
\begin{eqnarray}
y_{r}\partial_{y_{s}}-y_{m+s}\partial_{
y_{m+r}}+\kappa_{r}\partial_{\kappa_{s}}-\kappa_{m+s}\partial_{\kappa_{m+r}}
\end{eqnarray}
and
\begin{eqnarray}
y_{r}\partial_{y_{m+s}}-y_{s}\partial_{
y_{m+r}}+\kappa_{r}\partial_{\kappa_{m+s}}-\kappa_{s}\partial_{\kappa_{m+r}}
\end{eqnarray}
for $1\leq r<s\leq m$ (c.f. (2.14) and (2.57)). They annihilate
$v_{1}$. in the case of $n=2m+1$, we have additional  operators of
positive roots:
\begin{eqnarray}
y_{r}\partial_{y_{0}}-y_{0}\partial
y_{m+r}+\kappa_{r}\partial_{\kappa_{0}}-\kappa_{0}\partial
\kappa_{m+r}.
\end{eqnarray}
Then (2.84)-(2.86)  annihilate $v_{1}$.

Since $y_{1}^{k}$ is harmonic, $v_{1}$ is a singular vector of
${\mathcal {H}}'_kV_{\mbb{C}}$. The weight of $y_{1}^{k}$ is
$k\lambda_{1}$ if $n>4$, is $k(\lambda_1+\lambda_2)$ if $n=4$ and is
$2k\lambda_{1}$ if $n=3$. The weight of $\kappa_{1}$ is
$\lambda_{1}$ if $n>4$, is $\lambda_1+\lambda_2$ if $n=4$ and is
$2\lambda_{1}$ if $n=3$. Then the weight of $v_{1}$ is
$(k+1)\lambda_{1}$ if $n>4$, is $(k+1)(\lambda_1+\lambda_2)$if $n=4$
and is $2(k+1)\lambda_{1}$ if $n=3$, which implies that $v_1$ is the
highest weight vector of $W_1$. Furthermore,
$\Delta(v_{1})+b(\nabla^T\cdot\nabla)
(v_{1})=b(\nabla^T\cdot\nabla)(v_{1})$. But
\begin{eqnarray} \nabla
(v_{1})&=&\frac{1}{2}\partial_{x_{1}}(y_{1}^{k})+\frac{i}{2}\partial_{x_{m+1}}(y_{1}^{k})\nonumber\\&=&
\frac{k}{2^{k+1}}(x_{1}+ix_{m+1})^{k-1}+i^{2}\frac{k}{2^{k+1}}(x_{1}+ix_{m+1})^{k-1}=0,
\end{eqnarray}
which implies that $v_1$ is a solution of Navier equations. By the
invariant property of Navier equations under $o(n,\mbb{R})$,
$\hat{\cal H}_{k,1}$ is a solution subspace.

 (b) Now we approach the highest weight vectors in $W_{2}$. It is
 obviously
 that $v_2\in {\mathcal
{H}}_{k}'V_{\mbb{C}}$.

If $m>1$, then
\begin{eqnarray}
&&(E_{r,s}-E_{m+s,m+r})(v_2)\nonumber\\
&=&(y_{r}\partial_{y_{s}}-y_{m+s}\partial_{
y_{m+r}}+\kappa_{r}\partial_{\kappa_{s}}-\kappa_{m+s}\partial_{\kappa_{m+r}})
(-y_{2}y_{1}^{k-1}\kappa_{1}+y_{1}^{k}\kappa_{2})=0
\end{eqnarray} for $2\leq r< s\leq m$,
\begin{eqnarray}
&&(E_{1,2}-E_{m+2,m+1})(v_2)\nonumber\\
&=&(y_{1}\partial_{y_{2}}-y_{m+2}\partial_{
y_{m+1}}+\kappa_{1}\partial_{\kappa_{2}}-\kappa_{m+2}\partial_{
\kappa_{m+1}})(-y_{2}y_{1}^{k-1}\kappa_{1}+y_{1}^{k}\kappa_{2})\nonumber\\
&=&-y_{1}^{k}\kappa_{1}+y_{1}^{k}\kappa_{1}=0
\end{eqnarray} and
\begin{eqnarray}
&&(E_{r,m+s}-E_{s,m+r})(v_2)\nonumber\\
&=&(y_{r}\partial_{y_{m+s}}-y_{s}\partial_{
y_{m+r}}+\kappa_{r}\partial_{\kappa_{m+s}}-\kappa_{s}\partial_{\kappa_{m+r}})
(-y_{2}y_{1}^{k-1}\kappa_{1}+y_{1}^{k}\kappa_{2})=0.
\end{eqnarray}
Moreover, if $n=2m+1$ is odd,
\begin{eqnarray}
&&(E_{r,0}-E_{0,m+r})(v_2)\nonumber\\
&=&(y_{r}\partial_{y_{0}}-y_{0}\partial
y_{m+r}+\kappa_{r}\partial_{\kappa_{0}}-\kappa_{0}\partial
\kappa_{m+r})(-y_{2}y_{1}^{k-1}\kappa_{1}+y_{1}^{k}\kappa_{2})=0.
\end{eqnarray}
 For
$n=3$, $v_{2}=-y_{0}y_{1}^{k-1}\kappa_{1}+y_{1}^{k}\kappa_{0}$. The
operator of positive root vector of $o'(3,\mbb{C})$ is
\begin{eqnarray}
y_{1}\partial_{y_{0}}-y_{0}\partial_{
y_{2}}+\kappa_{1}\partial_{\kappa_{0}}-\kappa_{0}\partial_{\kappa_{2}}
\end{eqnarray}
which also annihilates $v_{2}$. in the case of $n=4$, we have that
\begin{eqnarray}
&&(y_{1}\partial_{y_{4}}-y_{2}\partial_{y_{3}}
+\kappa_{1}\partial_{\kappa_{4}}-\kappa_{2}\partial_{\kappa_{3}})(v_{2,+})\nonumber
\\&=&(y_{1}\partial_{y_{4}}+\kappa_{1}\partial_{\kappa_{4}})
(-y_{4}y_{1}^{k-1}\kappa_{1}+y_{1}^{k}\kappa_{4})=
-y_{1}^{k}\kappa_{1}+y_{1}^{k}\kappa_{1}=0
\end{eqnarray}
and similarly for $v_{2,-}$.

 Note that
the weight of $v_{2}$ is $(k+1)\lambda_{1}-\alpha_{1}$
($2(k+1)\lambda_{1}-\alpha_{1}$ if $n=3$). So it is the highest
weight of $W_2$. Moreover, the  weight of vector $v_{2,-}$ (res.
$v_{2,+}$) is $(k+1)(\lambda_{1}+\lambda_2)-\alpha_{1}$ (res.
$(k+1)(\lambda_{1}+\lambda_2)-\alpha_{2}$). Hence it is the highest
weight of $W_{2,-}$ (res. $W_{2,+}$).

In the case of $n\neq3$,
\begin{eqnarray}
2^{k+1}\nabla
(v_{2})&=&\partial_{x_{1}}(-(x_{2}+ix_{m+2})(x_{1}+ix_{m+1})^{k-1})
+\partial_{x_{2}}(x_{1}+ix_{m+1})^{k}\nonumber\\
&&+\partial_{x_{m+1}}(-i(x_{2}+ix_{m+2})(x_{1}+ix_{m+1})^{k-1})
+\partial_{x_{m+2}}(i(x_{1}+ix_{m+1})^{k})\nonumber\\
&=&-(k-1)(x_{2}+ix_{m+2})(x_{1}+ix_{m+1})^{k-2}\nonumber\\&&-
i^{2}(k-1)(x_{2}+ix_{m+2})(x_{1}+ix_{m+1})^{k-2} =0
\end{eqnarray}
and it is easy to verify  $\nabla(v_{2,-})=0$. Moreover,
\begin{eqnarray}
2^{k+1}\nabla
(v_{2,+})&=&\partial_{x_{1}}(-(ix_{2}+x_{4})(x_{1}+ix_{3})^{k-1})
+\partial_{x_{2}}(i(x_{1}+ix_{3})^{k})\nonumber\\
&&+\partial_{x_{3}}(-i(ix_{2}+x_{4})(x_{1}+ix_{3})^{k-1})
+\partial_{x_{4}}((x_{1}+ix_{3})^{k})=0.
\end{eqnarray}
in the case of $n=3$,
\begin{eqnarray}
2^{k+1}\nabla(v_{2})&=&\partial_{x_{0}}((x_{1}+ix_{2})^{k-1}(1+i))
+\partial_{x_{1}}(-(1+i)x_{0}(x_{1}+ix_{2})^{k-1})\nonumber\\
&&+\partial_{x_{2}}(-(1+i)ix_{0}(x_{1}+ix_{2})^{k-1})=0.
\end{eqnarray}
Since $y_{1}^{k}$, $y_{2}y_{1}^{k-1}$ and $y_{0}y_{1}^{k-1}$ are
harmonic, $\Delta(v_{2})+b(\nabla^T\cdot\nabla)
(v_{2})=b(\nabla^T\cdot\nabla)(v_{2})=0$. Similarly, the same
equation holds for $v_{2,\mp}$. That is, $v_2$ and $v_{2,\mp}$ are
solutions of Navier equations. Hence $\hat{\cal H}_{k,2}$ is a
solution subspace.

 (c) We now deal with the
highest weight vectors of $W'_{3}$.

 in the case of $n=2m$, observe that $E_{m-1,2m}-E_{m,2m-1}$ annihilate
 $v'_3$ and
\begin{eqnarray}
&&(y_{l}\partial_{y_{l+1}}-y_{m+l+1}\partial_{y_{m+l}}+
\kappa_{l}\partial_{\kappa_{l+1}}-\kappa_{m+l+1}\partial_{\kappa_{m+l}})(v_{3}^{'})\nonumber\\
&=&
2(k-1)y_{l}y_{m+l+1}y_{1}^{k-2}\kappa_{1}-(2k+n-4)y_{l}y_{1}^{k-1}\kappa_{m+l+1}
-2(k-1)y_{m+l+1}y_{l}y_{1}^{k-2}\kappa_{1}\nonumber\\
&&+(2k+n-4)y_{m+l+1}y_{1}^{k-1}\kappa_{l}
-(2k+n-4)y_{m+l+1}y_{1}^{k-1}\kappa_{l}+(2k+n-4)y_{l}y_{1}^{k-1}\kappa_{m+l+1}\nonumber\\
&=&0
\end{eqnarray} for $l=1,\ldots,m-1.$
On the other hand,
\begin{eqnarray}
&&\Delta(2(k-1)\sum\limits_{j=1}^{m}y_{j}y_{m+j}y_{1}^{k-2}-
(2k+n-4)y_{m+1}y_{1}y_{1}^{k-2})\nonumber\\
&=&i(2(k-1)(k-1))-(2k+n-4)(k-1)+2(k-1)(m-1))y_{1}^{k-2}=0,
\end{eqnarray} and
$\Delta y_{j}y_{1}^{k-2}=0$ for $j\neq m+1$ by (2.68). That is,
$v_{3}^{'}\in\mathcal {H}'_k V_{\mbb{C}}$ is harmonic , and so it is
a singular vector of $\mathcal {H}'_k V_{\mbb{C}}$ by (a), (b) and
Lemma 2.3.

in the case of $n=2m+1$, we similarly have that
\begin{eqnarray}
(y_{l}\partial_{y_{l+1}}-y_{m+l+1}\partial_{y_{m+l}}+
\kappa_{l}\partial_{\kappa_{l+1}}-\kappa_{m+l+1}\partial_{\kappa_{m+l}})(v_{3}^{'})
=0.
\end{eqnarray} and
\begin{eqnarray}
 &&(y_{r}\partial_{y_{0}}-y_{0}\partial_{y_{m+r}}+
 \kappa_{r}\partial_{\kappa_{0}}-\kappa_{0}\partial_{\kappa_{m+r}})
(v_{3}^{'})\nonumber\\&=&
-(2k+n-4)y_{r}y_{1}^{k-1}\kappa_{0}+2(k-1)y_{r}y_{0}y_{1}^{k-2}\kappa_{1}
-2(k-1)y_{r}y_{0}y_{1}^{k-2}\kappa_{1}\nonumber\\&&
+(2k+n-4)y_{0}y_{1}^{k-1}\kappa_{r}
-(2k+n-4)y_{0}y_{1}^{k-1}\kappa_{r}+(2k+n-4)y_{r}y_{1}^{k-1}\kappa_{0}\nonumber\\&=&0.
\end{eqnarray} In addition,
\begin{eqnarray}
&&\Delta(2(k-1)(\sum\limits_{j=1}^{m}y_{j}y_{m+j}+\frac{1}{2}y_{0}^{2})y_{1}^{k-2}-
(2k+n-4)y_{m+1}y_{1}y_{1}^{k-2})\nonumber\\
&=&\frac{i}{2}(k-1)(2n+4(k-2))y_{1}^{k-2}-i(2k+n-4)(k-1)y_{1}^{k-2}
=0
\end{eqnarray}
and $\Delta y_{j}y_{1}^{k-2}=0$ for $j\neq m+1$ by (2.68). That is,
$v_{3}^{'}\in\mathcal {H}'_k V_{\mbb{C}}$ is harmonic, and so it is
a singular vector of $\mathcal {H}'_k V_{\mbb{C}}$ by (a), (b) and
Lemma 2.3. Finally,  one get $v'_3\in W'_3$ by checking the weight
of $v'_3$. $\qquad\Box$
 \vspace{0.4cm}

 Set
\begin{equation}
v_{3}^{''}=2\sum\limits_{j=1}^{m}y_{j}y_{m+j}y_{1}^{k-2}\kappa_{1}\;\;
\mbox{if}\; n=2m
\end{equation} and
\begin{eqnarray}
v_{3}^{''}=(2\sum\limits_{j=1}^{m}y_{j}y_{m+j}+y_{0}^{2})y_{1}^{k-2}\kappa_{1}\;\;
\mbox{if}\; n=2m+1.
\end{eqnarray}
 Since
$[x_{r}\partial_{x_{s}}-x_{s}\partial_{x_{r}},\sum_{l=1}^{n}x_{l}^{2}]=0$,
the irreducible module with $v_3^{''}$ as its highest weight vector
is isomorphic to $V((k-1)\lambda_{1})$ for $n>4$,
$V((k-1)(\lambda_{1}+\lambda_{2}))$ for $n=4$ and
$V(2(k-1)\lambda_{1})$ for $n=3$. Let
\begin{eqnarray}
c=\frac{(2k+n-2)(k+n-3)(k-1)}{2(b^{-1}(2k+n-4)+k-1)}.
\end{eqnarray}
\begin{lemma} The vector
$v_{3}=v_{3}^{'}+cv_{3}^{''}$ is a complex solution of Navier
equations. Then the complex irreducible $o(n,\mathbb{C})$-module
$W_{3}$ generated by $v_{3}$ is a complex solution subspace of
Navier equations.
\end{lemma}

{\it Proof}.  Suppose $n=2m$. We have
\begin{eqnarray}
v_{3}^{'}&=&2(k-1)\sum\limits_{j=1}^{m}y_{j}y_{m+j}y_{1}^{k-2}\kappa_{1}
-(2k+n-4)\sum\limits_{j=1}^{m}y_{1}^{k-1}(y_{m+j}\kappa_{j}+y_{j}\kappa_{m+j})\nonumber\\
&=&
\frac{i}{2^{k}}(k-1)\sum\limits_{j=1}^{n}x_{j}^{2}(x_{1}+ix_{m+1})^{k-2}\vs_{1}-
\frac{1}{2^{k}}(k-1)\sum\limits_{j=1}^{n}x_{j}^{2}(x_{1}+ix_{m+1})^{k-2}\vs_{m+1}\nonumber\\&&
-i\frac{2k+n-4}{2^{k}}\sum\limits_{s=1}^{n}x_{s}(x_{1}+ix_{m+1})^{k-1}\vs_{s}
\end{eqnarray}
\begin{eqnarray}
v_{3}^{''}&=& \sum\limits_{j=1}^{m}y_{j}y_{m+j}y_{1}^{k-2}\vs_{1}+
i\sum\limits_{j=1}^{m}y_{j}y_{m+j}y_{1}^{k-2}\vs_{m+1}\nonumber\\&=&
\frac{i}{2^{k}}\sum\limits_{j=1}^{n}x_{j}^{2}(x_{1}+ix_{m+1})^{k-2}\vs_{1}
-\frac{1}{2^{k}}\sum\limits_{j=1}^{n}x_{j}^{2}(x_{1}+ix_{m+1})^{k-2}\vs_{m+1}.
\end{eqnarray}
Hence
\begin{eqnarray}
\Delta (v_{3}^{'})+b(\nabla^T\cdot\nabla)(v_{3}^{'})&=&
-\frac{i(k-1)}{2^{k}}b(k+n-3)(2k+n-2)(x_{1}+ix_{m+1})^{k-2}\vs_{1}\nonumber\\&&+
\frac{(k-1)}{2^{k}}b(k+n-3)(2k+n-2)(x_{1}+ix_{m+1})^{k-2}\vs_{m+1},\nonumber\\
\end{eqnarray}
\begin{eqnarray}
\Delta (v_{3}^{''})+b(\nabla^T\cdot\nabla)(v_{3}^{''})&=&
\frac{i}{2^{k}}(2(2k+n-4)+2b(k-1))(x_{1}+ix_{m+1})^{k-2}\vs_{1}\nonumber\\
&&-\frac{1}{2^{k}}(2(2k+n-4)+2b(k-1))(x_{1}+ix_{m+1})^{k-2}\vs_{m+1}.\nonumber\\
\end{eqnarray}

Assume $n=2m+1$. Then
\begin{eqnarray}
v_{3}^{'}&=&
    \frac{i}{2^{k}}(k-1)\sum\limits_{j=0}^{2m}x_{j}^{2}(x_{1}+ix_{m+1})^{k-2}\vs_{1}-
    \frac{1}{2^{k}}(k-1)\sum\limits_{j=0}^{2m}x_{j}^{2}(x_{1}+ix_{m+1})^{k-2}\vs_{m+1}
    \nonumber\\&&
    -i\frac{2k+n-4}{2^{k}}\sum\limits_{s=0}^{2m}x_{s}(x_{1}+ix_{m+1})^{k-1}\vs_{s},
\end{eqnarray}
\begin{eqnarray}
v_{3}^{''}=\frac{i}{2^{k}}\sum\limits_{j=0}^{2m}x_{j}^{2}(x_{1}+ix_{m+1})^{k-2}\vs_{1}
-\frac{1}{2^{k}}\sum\limits_{j=0}^{2m}x_{j}^{2}(x_{1}+ix_{m+1})^{k-2}\vs_{m+1}.
\end{eqnarray}
Thus \begin{eqnarray} \Delta(v_{3}^{'})+b(\nabla^T\cdot\nabla)
(v_{3}^{'})&=&
-\frac{i(k-1)}{2^{k}}b(k+n-3)(2k+n-2)(x_{1}+ix_{m+1})^{k-2}\vs_{1}\nonumber\\&&+
\frac{(k-1)}{2^{k}}b(k+n-3)(2k+n-2)(x_{1}+ix_{m+1})^{k-2}\vs_{m+1},\nonumber\\
\end{eqnarray}
\begin{eqnarray}
\Delta (v_{3}^{''})+b(\nabla^T\cdot\nabla )(v_{3}^{''})&=&
\frac{i}{2^{k}}(2(2k+n-4)+2b(k-1))(x_{1}+ix_{m+1})^{k-2}\vs_{1}\nonumber\\
&&-\frac{1}{2^{k}}(2(2k+n-4)+2b(k-1))(x_{1}+ix_{m+1})^{k-2}\vs_{m+1}.\nonumber\\
\end{eqnarray}
The conclusion follows from (2.107), (2.108) and (2.111), (2.112).
 \hspace{1cm}$\Box$ \vspace{0.4cm}

 Now we can describe the
irreducible modules of $o(n,\mbb{R})$ which are solution spaces of
Navier equations.
\begin{lemma} The linear map
determined by
\begin{equation}
\psi(x_{i_{1}}\cdots
x_{i_{k+1}})=\sum\limits_{j=1}^{k+1}x_{i_{1}}\cdots
x_{i_{j-1}}x_{i_{j+1}}\cdots x_{i_{k+1}}\varsigma_{i_{j}}
\end{equation}
is an $o(n,\mbb{R})$-module isomorphism from $ {\cal H}_{k+1}$ to
$\hat{\cal H}_{k,1}$.
\end{lemma}

{\it Proof}. Set $f=x_{1}^{l_{1}}\cdots x_{n}^{l_{n}}$, and then
$x_{1}\partial_{x_{2}}(f)=l_{2}x_{1}^{l_{1}+1}x_{2}^{l_{2}-1}x_{3}^{l_{3}}\cdots
x_{n}^{l_{n}}$. It follows that
\begin{eqnarray}
&&\psi(x_{1}\partial_{x_{2}}(f))\nonumber\\\nonumber
&=&\psi(l_{2}x_{1}^{l_{1}+1}x_{2}^{l_{2}-1}x_{3}^{l_{3}}\cdots
x_{n}^{l_{n}})\\\nonumber &=&
l_{2}(l_{1}+1)x_{1}^{l_{1}}x_{2}^{l_{2}-1}x_{3}^{l_{3}}\cdots
x_{n}^{l_{n}}\varsigma_{1}+
l_{2}(l_{2}-1)x_{1}^{l_{1}+1}x_{2}^{l_{2}-2}x_{3}^{l_{3}}\cdots
x_{n}^{l_{n}}\varsigma_{2}\\  &&+
l_{2}\sum\limits_{j=3}^{n}l_jx_{1}^{l_{1}+1}x_{2}^{l_{2}-1}x_{3}^{l_{3}}\cdots
x_{j-1}^{l_{j-1}}x_{j}^{l_{j}-1}x_{j+1}^{l_{j+1}}\cdots
x_{n}^{l_{n}}\varsigma_{j}.
\end{eqnarray}
On the other hand,
$\psi(f)=\sum\limits_{j=1}^{n}l_{j}x_{1}^{l_{1}}\cdots
x_{j-1}^{l_{j-1}}x_{j}^{l_{j}-1}x_{j+1}^{l_{j+1}}\cdots
x_{n}^{l_{n}}\varsigma_{j}$. Then
\begin{eqnarray}
(x_{1}\partial_{x_{2}}+\varsigma_{1}\partial_{\varsigma_{2}})(f)&=&
(x_{1}\partial_{x_{2}}+\varsigma_{1}\partial_{\varsigma_{2}})
(\sum\limits_{j=1}^{n}l_{j}x_{1}^{l_{1}}\cdots
x_{j-1}^{l_{j-1}}x_{j}^{l_{j}-1}x_{j+1}^{l_{j+1}}\cdots
x_{n}^{l_{n}}\varsigma_{j})\nonumber\\&=&
l_{1}l_{2}x_{1}^{l_{1}}x_{2}^{l_{2}-1}x_{3}^{l_{3}}\cdots
x_{n}^{l_{n}}\varsigma_{1}+l_{2}(l_{2}-1)x_{1}^{l_{1}+1}x_{2}^{l_{2}-2}x_{3}^{l_{3}}\cdots
x_{n}^{l_{n}}\varsigma_{2}\nonumber\\&&+
l_{2}\sum\limits_{j=3}^{n}l_jx_{1}^{l_{1}+1}x_{2}^{l_{2}-1}x_{3}^{l_{3}}\cdots
x_{j-1}^{l_{j-1}}x_{j}^{l_{j}-1}x_{j+1}^{l_{j+1}}\cdots
x_{n}^{l_{n}}\varsigma_{j}\nonumber\\&&+
l_{2}x_{1}^{l_{1}}x_{2}^{l_{2}-1}x_{3}^{l_{3}}\cdots
x_{n}^{l_{n}}\varsigma_{1}\nonumber\\
&=&\psi(x_{1}\partial_{x_{2}}(f)).
\end{eqnarray}
By symmetry, $\psi(A(f))=A(\psi(f))$ for any $A\in gl(n,\mbb{R})$.
Moveover, $f$ is harmonic and
$\psi(f)=\sum\limits_{j=1}^{n}g_{j}\varsigma_{j}$, then all $g_{j}$
are harmonic. Indeed, we write
$f=\sum\limits_{l=0}^{k+1}f_{l}(x_{2},\ldots,x_{n})\frac{x_{1}^{l}}{l!},$
and then $f$ is harmonic if and only if  $f_{l+2}=-\Delta f_{l}$ for
$0\leq l\leq k-1$. It follows that
$g_{1}=\sum\limits_{l=0}^{k}f_{l+1}(x_{2},\ldots,x_{n})\frac{x_{1}^{l}}{l!}$
is harmonic. By symmetry, all $g_j$ are harmonic. Hence
$\psi(\mathcal {H}_{k+1})\subset\mathcal {H}_{k} V$  is an
irreducible module of $o(n,\mbb{R})$. Thus $\psi(\mathcal
{H}_{k+1})=\hat{\cal H}_{k,1}$ by Lemma 2.3 and Lemma 2.4.
\hspace{1cm}$\Box$ \vspace{0.4cm}

\begin{lemma} i) Denote $\tilde{x}_{j}=(k-1)\sum\limits_{r=1}^{n}x_{r}^{2}\varsigma_{j}-
(2k+n-4)x_{j}\sum\limits_{r=1}^{n}x_{r}\varsigma_{r}$ for $1\leq
j\leq n$. Then the linear map $\varphi_1$ determined by
\begin{equation}
x_{i_1}\cdots x_{i_{k-1}}\longmapsto
\sum\limits_{s=1}^{k-1}x_{i_1}\cdots x_{i_{s-1}}x_{i_{s+1}}\cdots
x_{i_{k-1}}\tilde{x}_{i_{s}}
\end{equation}
is an $o(n,\mbb{R})$-module isomorphism from $\mathcal {H}_{k-1}$ to
$\hat{\cal H}_{k,3}^{'}$. Moreover,
$\varphi_{2}=(\sum\limits_{r=1}^{n}x_{r}^{2})\psi$ is an
$o(n,\mbb{R})$-module isomorphism from $\mathcal {H}_{k-1}$ to the
real form $\hat{\cal H}_{k,3}^{''}$ of the complex
$o'(n,\mbb{R})$-submodule generated by ${v}_{3}^{''}$ (cf. (2.102)
and (2.103)).

ii) The module $W_3$ is of real type,
$\varphi=\varphi_{1}+c\varphi_{2}$ (cf. (2.104)) is an
$o(n,\mbb{R})$-module isomorphism from $\mathcal {H}_{k-1}$ to the
real form (denoted by $\hat{\cal H}_{k,3}$) of $W_3$  (cf. Lemma
2.6).
\end{lemma}
{\it Proof}. Firstly, it is clear that
 \begin{eqnarray}
 (x_{1}\partial_{x_{2}}-x_{2}\partial_{x_{1}}+ \varsigma_{1}\partial_{ \varsigma_{2}}- \varsigma_{2}\partial_{ \varsigma_{1}})
(\tilde{x}_{1})&=&-\tilde{x}_{2},\\
(x_{1}\partial_{x_{2}}-x_{2}\partial_{x_{1}}+
\varsigma_{1}\partial_{ \varsigma_{2}}- \varsigma_{2}\partial_{
\varsigma_{1}}) (\tilde{x}_{2})&=&\tilde{x}_{1},\\
(x_{1}\partial_{x_{2}}-x_{2}\partial_{x_{1}}+
\varsigma_{1}\partial_{ \varsigma_{2}}- \varsigma_{2}\partial_{
\varsigma_{1}}) (\tilde{x}_{j})&=&0
\end{eqnarray} for $j>2$. Moreover,
\begin{eqnarray}\nonumber
\varphi_{1}((E_{1,2}-E_{2,1})(\zeta))&=&
l_{2}(l_{1}+1)\frac{\zeta}{x_{2}}\tilde{x}_{1}+
l_{2}(l_{2}-1)\frac{x_{1}\zeta}{x_{2}^{2}}\tilde{x}_{2}+
\sum\limits_{j=3}^{n}l_{2}l_{j}\frac{x_{1}\zeta}{x_{2}x_{j}}\tilde{x}_{j}\\
&&-l_{1}(l_{1}-1)\frac{x_{2}\zeta}{x_{1}^{2}}\tilde{x}_{1}-
l_{1}(l_{2}+1)\frac{\zeta}{x_{1}}\tilde{x}_{2}-
\sum\limits_{j=3}^{n}l_{1}l_{j}\frac{x_{2}\zeta}{x_{1}x_{j}}\tilde{x}_{j},\nonumber\\
\end{eqnarray}
where $\zeta=x_{1}^{l_{1}}\cdots x_{n}^{l_{n}}$. On the other hand,
\begin{eqnarray}
\nonumber&& (E_{1,2}-E_{2,1})(\varphi_{1}(\zeta))\\\nonumber&=&
l_{1}l_{2}\frac{\zeta}{x_{2}}\tilde{x}_{1}-
l_{1}(l_{1}-1)\frac{x_{2}\zeta}{x_{1}^{2}}\tilde{x}_{1}+
l_{1}\frac{\zeta}{x_{1}}((x_{1}\partial_{x_{2}}-x_{2}\partial_{x_{1}}+
\varsigma_{1}\partial_{ \varsigma_{2}}- \varsigma_{2}\partial_{
\varsigma_{1}})( \tilde{x}_{1}))\\\nonumber &&+
l_{2}(l_{2}-1)\frac{x_{1}\zeta}{x_{2}^{2}}\tilde{x}_{2}-
l_{1}l_{2}\frac{\zeta}{x_{1}}\tilde{x}_{2}+
l_{2}\frac{\zeta}{x_{2}}((x_{1}\partial_{x_{2}}-x_{2}\partial_{x_{1}}+
\varsigma_{1}\partial_{ \varsigma_{2}}- \varsigma_{2}\partial_{
\varsigma_{1}}) (\tilde{x}_{2}))\\\nonumber&&+
\sum\limits_{j=3}^{n}(l_{2}l_{j}\frac{x_{1}\zeta}{x_{2}x_{j}}\tilde{x}_{j}-
l_{1}l_{j}\frac{x_{2}\zeta}{x_{1}x_{j}}\tilde{x}_{j})+\sum\limits_{j=3}^{n}
\frac{\zeta}{x_{j}}((x_{1}\partial_{x_{2}}-x_{2}\partial_{x_{1}}+
\varsigma_{1}\partial_{ \varsigma_{2}}- \varsigma_{2}\partial_{
\varsigma_{1}})
(\td x_{j}))\\
\end{eqnarray}
by (2.117)-(2.119). Thus
$\varphi_{1}((E_{1,2}-E_{2,1})(\zeta))=(E_{1,2}-E_{2,1})(\varphi_{1}(\zeta))$.
By the symmetry on sub-indices, $\varphi_{1}$ is an
$o(n,\mbb{R})$-module monomorphism from $\mathcal {H}_{k-1}$ to
$\hat{\cal A}_k$. If
 $n=2m$,
\begin{eqnarray}
 &&\varphi_{1}(y_{1}^{k-1})\nonumber\\
 &=&\varphi_{1}(\frac{1}{2^{k-1}}(x_{1}+ix_{m+1})^{k-1})
 \nonumber\\
 &=&\frac{1}{2^{k-1}}\sum\limits_{r=0}^{k-1}{k-1\choose r}((k-1-r)
 x_{1}^{k-2-r}(ix_{m+1})^{r}\tilde{x}_{1}+rix_{1}^{k-1-r}(ix
_{m+1})^{r-1}\tilde{x}_{m+1})\nonumber\\
&=& \frac{1}{2^{k-1}}\sum\limits_{r=0}^{k-2}{k-2\choose
r}(k-1)x_{1}^{k-2-r}(ix
_{m+1})^{r}(\tilde{x}_{1}+i\tilde x_{m+1})\nonumber\\
&=& \frac{k-1}{2}y_{1}^{k-2}(\tilde{x}_{1}+i\tilde
x_{m+1})\nonumber\\&=& -2i(k-1)v_{3}^{'}
 \end{eqnarray}
(cf. (2.82)). So $\varphi_{1}$ is an $o(n,\mbb{R})$-module
 isomorphism from $\mathcal {H}_{k-1}$ to
$\hat{\cal H}_{k,3}^{'}$. It holds similarly when $n=2m+1$.

Lemma 2.8 implies that $\varphi_{2}$ is an $o(n,\mbb{R})$-module
monomorphism from $\mathcal {H}_{k-1}$ to $\hat{\cal A}_k$. If
 $n=2m$, we denote $\xi=\sum\limits_{j=1}^{n}x_{j}^{2}$,
 and  then
\begin{eqnarray}
&&\varphi_{2}(y_{1}^{k-1})\nonumber\\&=&
 \frac{1}{2^{k-1}}\sum\limits_{r=0}^{k-1}{k-1\choose r}
 ((k-1-r)\xi  x_{1}^{k-2-r}(ix_{m+1})^{r}\varsigma_{1}+ ri\xi
x_{1}^{k-1-r}(ix_{m+1})^{r-1} \varsigma_{m+1})\nonumber\\&=&
-4i(k-1)\sum\limits_{j=1}^{m}y_{j}y_{m+j}y_{1}^{k-2}\kappa_{1}=
-2i(k-1)v_{3}^{''}
\end{eqnarray}
(cf. (2.102)). So $\varphi_{2}=(\sum\limits_{r=1}^{n}x_{r}^{2})\psi$
is an $o(n,\mbb{R})$-module isomorphism from $\mathcal {H}_{k-1}$ to
the real form $\hat{\cal H}_{k,3}^{''}$.  It holds similarly when
$n=2m+1$. Since $\varphi(\mathcal {H}'_{k-1})=W_3$ and
$\varphi(\bar{f})=\overline{\varphi(f)}$ for $f\in \mathcal
{H}'_{k-1}$, we know that $W_3$ is of real type by Lemma 2.4.
Moreover, $\varphi(\mathcal {H}_{k-1})$ is the real form of $W_3$.
That is $\varphi(\mathcal {H}_{k-1})=\hat{\cal H}_{k,3}.$
$\qquad\Box$ \vspace{0.4cm}

\begin{lemma} The subspace
\begin{equation} \hat{\cal H}_{k,2}=\{\sum\limits_{r=1}^{n}
f_{r}\varsigma_{r}\ |\ \ f_{r}\in \mathcal {H}_{k},\; \  \
\sum\limits_{r=1}^{n}x_{r}f_{r}=0\}.
\end{equation}
\end{lemma}

{\it Proof}. Denote by $\tilde{V}$ the right side of above equality.
Note that the linear map $\nu: \sum_{r=1}^{n}
f_r\vs_r\longmapsto\sum_{r=1}^{n}x_rf_r $ is an
$o'(n,\mbb{C})$-module homomorphism from $\hat{\cal A}_{\mbb{C}}$ to
${\cal A}_{\mbb{C}}$ by Lemma 2.1. In particular,
\begin{equation}\nu(\kappa_r)=y_r\qquad\mbox{for}\;\;1\leq r\leq n.\end{equation} Moreover,
$\nu(\mbox{Re}(v_1))=\mbox{Re}(\nu(v_1))=\mbox{Re}(y_1^{k+1})\in\mathcal
{H}_{k+1}$ by (2.78) and the classical harmonic analysis. Lemma
2.1 also tells us that $\nu_{\hat{\cal A}}$ is an
$o(n,\mbb{R})$-module homomorphism from $\hat{\cal A}$ to ${\cal
A}$. Recall that $\hat{\mathcal H}_{k,1}$ is an irreducible
$o(n,\mbb{R})$-submodule generated by $\mbox{Re}(v_1)$ and ${\cal
H}_{k+1}$ is an irreducible $o(n,\mbb{R})$-submodule. So
$\nu|_{\hat{\mathcal H}_{k,1}}$ is an $o(n,\mbb{R})$-module
isomorphism from $\hat{\mathcal H}_{k,1}$ to ${\cal H}_{k+1}$.

According to (2.61) and (2.63),
\begin{equation}4\sum_{r=1}^my_ry_{m+r}=i\sum_{s=1}^nx_s^2,\qquad 2y_0^2=ix_0^2. \end{equation}
Thus
\begin{equation}\nu(v_3')=\frac{3-n-k}{2^k}(x_1^2+\cdots+x_n^2)i(x_1+ix_{m+1})^{k-1}\end{equation}
by (2.82) if $n=2m$. When $n=2m+1$, we shift index $r\mapsto r+1$
and have
\begin{equation}\nu(v_3')=\frac{3-n-k}{2^k}(x_1^2+\cdots+x_n^2)i(x_2+ix_{m+2})^{k-1}\end{equation}
by (2.83).  Thus $0\neq \nu(\mbox{Re}(v_3'))=\mbox{Re}(\nu(v_3'))\in
(x_1^2+\cdots+x_n^2){\cal H}_{k-1}$. By the
$o(n,\mbb{R})$-irreducibility of $\hat{\cal H}_{k,3}'$ and
$(x_1^2+\cdots+x_n^2){\cal H}_{k-1}$, $\nu|_{\hat{\cal H}_{k,3}'}$
is an $o(n,\mbb{R})$-module isomorphism from $\hat{\mathcal
H}'_{k,3}$ to $(x_1^2+\cdots+x_n^2){\cal H}_{k-1}$.

By (2.79)-(2.81) and (2.125), $\nu(v_2)=0$ and $\nu(v_{2,\pm})=0$.
Similarly, we have $\nu|_{\hat{\cal H}_{k,2}}=0$. Since ${\cal
H}_kV=\hat{\cal H}_{k,1}\oplus \hat{\cal H}_{k,2}\oplus \hat{\cal
H}_{k,3}$, $\mbox{ker}\:\nu|_{{\cal H}_kV}= \hat{\cal H}_{k,2}$,
that is, (2.124) holds.$\qquad\Box$ \vspace{0.4cm}

\begin{corollary} The subspace
\begin{equation}
\hat{\cal H}_{k,1}+\hat{\cal
H}_{k,2}=\{\sum\limits_{r=1}^{n}f_{r}\vs_{r}\ |\ f_{r}\in \mathcal
{H}_{k}, \  \ \sum\limits_{r=1}^{n}\partial_{x_{r}}(f_{r})=0\}.
\end{equation}
\end{corollary}
{\it Proof}. Denote by $\tilde{V}$ the set in right side in (2.129).
 Note
$(\ptl_{x_1}+i\ptl_{x_{m+1}})((x_1+ix_{m+1})^k)=0$. By (2.78),
$\mbox{Re}(v_1)\in\tilde{V}\bigcap\hat{\cal H}_{k,1}$. Moreover, the
map $\sigma:\sum\limits_{r=1}^{n}f_{r}\vs_{r}\rightarrow
\sum\limits_{r=1}^{n}\ptl_{x_r}(f_{r})$ is an $o(n,\mbb{R})$-module
homomorphism from $\hat{\cal A}$ to ${\cal A}$. The irreducibility
of $\hat{\cal H}_{k,1}$ as an $o(n,\mbb{R})$-submodule implies
$\hat{\cal H}_{k,1}\subset \tilde{V}$. For $\vec f\in \hat{\cal
H}_{k,2}$,  $0=\Delta(\sum\limits_{r=1}^{n}x_{r}f_{r})=
2\sum\limits_{r=1}^{n}\partial_{x_{r}}(f_{r})$ by straightforward
calculation. Thus $\hat{\cal H}_{k,2}\subset \tilde{V}$. Now
$\sgm(\hat{\cal H}_{k,3})=\sigma({\cal H}_kV)\neq\{0\}$. Since
$\hat{\cal H}_{k,3}$ is an irreducible $o(n,\mbb{R})$-submodule,
$\mbox{ker}\:\sigma|_{\hat{\cal H}_{3,k}}=\{0\}$. Hence (2.129)
holds. $\qquad\Box$\hspace{0.4cm}

Denote $d_{r,s}=x_{s}\partial_{x_{r}}-x_{r}\partial_{x_{s}}$ and
\begin{eqnarray}
\cal D=\left(\begin{array}{cccc}
                       0 & d_{3,4}&d_{4,2}&d_{2,3}\\
                       d_{4,3}& 0&d_{1,4}&d_{3,1}\\
                       d_{2,4}&d_{4,1}&0&d_{1,2}\\
                       d_{3,2}&d_{1,3}&d_{2,1}&0
                       \end{array}\right ).
\end{eqnarray}Then if $n=4,$ we have that
\begin{lemma}The subspaces
\begin{eqnarray}
\hat{\cal H}_{k,2\pm}=\{\vec{f}\in \hat{\cal H}_{k,2}|\ {\cal
D}\vec{f}=\pm(k+1)\vec{f}\ \}.
\end{eqnarray}
\end{lemma}
{\it Proof}. We can verify that $\mathcal {D}$  commutes  with
$o(4,\mathbb{R})$. Moreover, ${\cal D}(v_{2,\pm})=\pm(k+1)v_{2,\pm}$
by (2.80). Expression (2.131) holds because $v_{2,\pm}$ are
generators of  $\hat{\cal H}_{k,2\pm}$ as
$o(4,\mathbb{R})$-modules.$\qquad\Box$\vspace{0.4cm}

 Now we can get the main theorem of this section.
\begin{theorem}
Assume that integer  $n\geq 3$. Let $\hat{\cal H}_{k,1}=\psi({\cal
H}_{k+1})$ (cf. (2.113)) and $\hat{\cal H}_{k,3}=\varphi({\cal
H}_{k-1})$ (cf. Lemma 2.8), which are irreducible
$o(n,\mbb{R})$-submodules. Take $\hat{\cal H}_{k,2}$ in (2.124),
which is an irreducible $o(n,\mbb{R})$-submodule if $n\neq 4$. When
$n=4$, $\hat{\cal H}_{k,2}=\hat{\cal H}_{k,2+}\oplus \hat{\cal
H}_{k,2-}$ and $\hat{\cal H}_{k,2\pm}$ are irreducible
$o(4,\mbb{R})$-submodules characterized by (2.131). Then the
subspace of homogeneous polynomial solutions with degree $k$ of
Navier equation is $\hat{\cal H}_k=\hat{\cal H}_{k,1}\oplus
\hat{\cal H}_{k,2}\oplus \hat{\cal H}_{k,3}$, and
\begin{equation}\hat{\cal A}_k=\hat{\cal H}_k\oplus
(x_1^2+\cdots+x_n^2)\hat{\cal A}_{k-2}.\end{equation}
\end{theorem}
{\it Proof}. We have $\hat{\cal H}_{k,j}\subset\hat{\mathcal
{H}}_{k}$ for $j=1,2,3$ by Lemmas 2.5 and 2.6. By Xu's method, we
can calculate $\dim\mathcal {\hat{H}}_{k}=n\dim\mathcal {H}_{k}=\dim
\hat{\cal H}_{k,1}+\dim \hat{\cal H}_{k,2}+\dim \hat{\cal H}_{k,3}$
(The details will be given later for technical convenience (see
Proposition 3.8)). Since
\begin{equation}\hat{\cal H}_k\equiv {\cal
H}_kV\;\mbox{mod}\;(x_1^2+\cdots+x_n^2){\cal H}_{k-2}V\end{equation}
by the way of our taking subspaces $\hat{\cal H}_{k,1}, \hat{\cal
H}_{k,2}$ and $\hat{\cal H}_{k,3}$ in the theorem, (2.132) follows
from the facts ${\cal A}_k={\cal H}_k+(x_1^2+\cdots+x_n^2){\cal
H}_{k-2},\;\hat{\cal A}_k={\cal A}_kV$ and induction on $k$.
$\qquad\Box$

\section{Bases}
In this section, we will construct some bases of the subspaces
$\hat{\cal H}_{k,1}$, $\hat{\cal H}_{k,2}$, $\hat{\cal H}_{k,3}$ and
$\hat{\cal H}_{k,2\mp}$ defined in section 2.

Since $\mathcal {H}_{k+1}\stackrel{\psi}{\cong} \hat{\cal H}_{k,1}$
(cf. Lemma 2.7) and $\mathcal {H}_{k-1}\stackrel{\varphi}{\cong}
\hat{\cal H}_{k,3}$ (cf. Lemma 2.8), we can obtain the bases of
$\hat{\cal H}_{k,1}$ and $\hat{\cal H}_{k,3}$ by a basis of
$\mathcal {H}_{k}$ introduced in [X1]:

\begin{equation}\{w(\es,l_2,\ldots,l_n)\mid
\es\in\{0,1\};\;l_2,...,l_n\in\mbb{N},\;\es+\sum_{j=2}^nl_j=k\}
\end{equation}
with
\begin{equation}
w(\es,l_2,\ldots,l_n)=\sum_{r_2,...,r_n=0}^{\infty}
\frac{(-1)^{\sum\limits_{j=2}^{n}r_{j}} {r_2+\cdots+r_n\choose
r_2,...,r_n}\prod\limits_{s=2}^n{l_s\choose
2r_s}}{(1+2\es\sum\limits_{j=2}^{n}r_{j}){2(r_2+\cdots+r_n)\choose
2r_2,...,2r_n}}x_1^{\es+2\sum\limits_{j=2}^{n}r_{j}}
\prod\limits_{j=2}^{n}x_j^{l_j-2r_j}.
 \end{equation}
Take $\es\in\{0,1\}$, $l_2,...,l_n\in\mbb{N}$ such that
$\es+\sum_{j=2}^nl_j=k+1$, and define
\begin{eqnarray}
\vec{f}(\es,l_2,\ldots,l_n)=\psi(w(\es,l_2,\ldots,l_n))
=\sum_{j=1}^nf_j(\es,l_2,\ldots,l_n)\vs_j.
\end{eqnarray}
Then straightforward calculation shows that
\begin{eqnarray}
&&f_{1}(\epsilon,l_2,\ldots,l_n)\nonumber\\&=&\sum\limits_{r_{2},\ldots,r_{n}=0}^{\infty}
(\epsilon+2\sum\limits_{j=2}^{n}r_{j})\frac{(-1)^{\sum\limits_{j=2}^{n}r_{j}}
{r_{2}+\cdots+r_{n}\choose
r_{2},\cdots,r_{n}}\prod\limits_{s=2}^n{l_s\choose 2r_s}}
{(1+2\epsilon\sum\limits_{j=2}^{n}r_{j}){2(r_{2}+\cdots+r_{n})\choose
2r_{2},\cdots,2r_{n} }}
x_{1}^{\epsilon+2\sum\limits_{j=2}^{n}r_{j}-1}\prod\limits_{s=2}^{n}x_{s}^{l_{s}-2r_{s}}
\end{eqnarray} and for $j=2,\ldots,n$,
\begin{eqnarray}
&&f_{j}(\epsilon,l_2,\ldots,l_n)\nonumber\\&=&
\sum\limits_{r_{2},\ldots,r_{n}=0}^{\infty}(l_{j}-2r_{j})
\frac{(-1)^{\sum\limits_{j=2}^nr_{j}} {r_{2}+\cdots+r_{n}\choose
r_{2},\cdots,r_{n}}\prod\limits_{s=2}^n{l_s\choose 2r_s}}
{(1+2\epsilon\sum\limits_{j=2}^{n}r_{j}){2(r_{2}+\cdots+r_{n})\choose
2r_{2},\cdots,2r_{n} }} x_{1}^{\epsilon+2\sum\limits_{j=2}^{n}r_{j}}
x_j^{-1}\prod\limits_{s=2}^{n}x_{s}^{l_{s}-2r_{s}}.
\end{eqnarray}

\begin{proposition}
The set
\begin{eqnarray}
\{\vec{f}(\es,l_2,\ldots,l_n)|\epsilon=0\ \mbox{or}\
1;\;l_{j}\in\mathbb{N};\;\epsilon+\sum\limits_{j=2}^{n}l_{j}=k+1\ \}
\end{eqnarray} forms a basis of $\hat{\cal H}_{k,1}$, where
the components of $\vec{f}(\es,l_2,\ldots,l_n)$ are given by (3.4)
and (3.5).
\end{proposition}
Similarly, we define
\begin{equation}
\tilde{w}(\es,l_2,\ldots,l_n)=\sum_{r_2,...,r_n=0}^{\infty}
\frac{(-1)^{\sum\limits_{j=2}^{n}r_{j}} {r_2+\cdots+r_n\choose
r_2,...,r_n}\prod\limits_{s=2}^n{l_s\choose
2r_s}}{(1+2\es\sum\limits_{j=2}^{n}r_{j}){2(r_2+\cdots+r_n)\choose
2r_2,...,2r_n}}x_1^{\es+2\sum\limits_{j=2}^{n}r_{j}}
\prod\limits_{j=2}^{n}x_j^{l_j-2r_j},
 \end{equation}
where $\es\in\{0,1\}$, $l_2,...,l_n\in\mbb{N}$ and
$\es+\sum_{j=2}^nl_j=k-1$. Thus $\tilde{w}\in\mathcal {H}_{k-1}$.
 Denote
\begin{eqnarray}
\vec{g}(\es,l_2,\ldots,l_n)=\varphi(\tilde{w}(\es,l_2,\ldots,l_n))
=\sum_{j=1}^ng_j(\es,l_2,\ldots,l_n)\vs_j,
\end{eqnarray}
 then we get that
\begin{eqnarray}\nonumber
\nonumber&&g_{1}(\epsilon,l_2,\ldots,l_n)=
\sum\limits_{r_{2},\ldots,r_{n}=0}^{\infty}
\frac{(-1)^{\sum\limits_{j=2}^{n}r_{j}} {r_{2}+\cdots+r_{n}\choose
r_{2},\cdots,r_{n}}\prod\limits_{s=2}^n{l_s\choose 2r_s}}
{(1+2\epsilon\sum\limits_{j=2}^{n}r_{j}){2(r_{2}+\cdots+r_{n})\choose
2r_{2},\cdots,2r_{n} }}((\epsilon+2\sum\limits_{j=2}^{n}r_{j})
\\\nonumber&\times&((k-1)+\frac{(2k+n-2)(k+n-3)(k-1)}{2(b^{-1}(2k+n-4)+k-1)})
 (\sum\limits_{p=1}^{n}x_{p}^{2})
x_{1}^{\epsilon+2\sum\limits_{j=2}^{n}r_{j}-1}\prod\limits_{s=2}^{n}x_{s}^{l_{s}-2r_{s}}
\\&-&(k-1)(2k+n-4)x_{1}^{\epsilon+2\sum\limits_{j=2}^{n}r_{j}+1}
\prod\limits_{s=2}^{n}x_{s}^{l_{s}-2r_{s}})
\end{eqnarray}
and for $j=2,\ldots,n$,
\begin{eqnarray}\nonumber
\nonumber&&g_{j}(\epsilon,l_2,\ldots,l_n)=
\sum\limits_{r_{2},\ldots,r_{n}=0}^{\infty}\frac{(-1)^{\sum\limits_{j=2}^{n}r_{j}}
{r_{2}+\cdots+r_{n}\choose
r_{2},\cdots,r_{n}}\prod\limits_{s=2}^n{l_s\choose 2r_s}}
{(1+2\epsilon\sum\limits_{j=2}^{n}r_{j}){2(r_{2}+\cdots+r_{n})\choose
2r_{2},\cdots,2r_{n} }} ((l_{j}-2r_{j})\\\nonumber&\times&
((k-1)+\frac{(2k+n-2)(k+n-3)(k-1)}{2(b^{-1}(2k+n-4)+k-1)})
 (\sum\limits_{p=1}^{n}x_{p}^{2})
x_{1}^{\epsilon+2\sum\limits_{j=2}^{n}r_{j}}
x_j^{-1}\prod\limits_{s=2}^{n}x_{s}^{l_{s}-2r_{s}}
\\&-&(k-1)(2k+n-4)x_{1}^{\epsilon+2\sum\limits_{j=2}^{n}r_{j}}x_{j}
\prod\limits_{s=2}^{n}x_{s}^{l_{s}-2r_{s}}).
\end{eqnarray}

\begin{proposition}
The set
\begin{eqnarray}
\{\vec{g}(\es,l_2,\ldots,l_n)|\epsilon=0\ \mbox{or}\
1,\;l_{j}\in\mathbb{N},\;\epsilon+\sum\limits_{j=2}^{n}l_{j}=k-1\ \}
\end{eqnarray} forms a basis of $\hat{\cal H}_{k,3}$, where
the components of $\vec{g}(\es,l_2,\ldots,l_n)$ are given by (3.9)
and (3.10).
\end{proposition}\vspace{0.4cm}

Now we will find a basis of $\hat{\cal{H}}_{k,2}$ by solving the
equation in (2.124). Let
$\vec{f}=\sum\limits_{j=1}^nf_j\vs_j\in\hat{\cal{H}}_{k,2}$. Then
$f_j\in {\cal{H}}_k$ for $j=1,\ldots,n$ and
$f_n=-x_n^{-1}\sum\limits_{j=1}^{n-1}x_jf_j$. We write
\begin{eqnarray}
f_{j}=\sum\limits_{l=0}^{k}f_{n-1,l}^{j}(x_{1},\ldots,x_{n-1})\frac{x_{n}^{l}}{l!}
\;\;\;\;\mbox{for}\;\; j=1,\ldots,n,
\end{eqnarray}
where $f_{n-1,l}^{j}(x_{1},\ldots,x_{n-1})$ are homogeneous
polynomials with degree $k-l$. Denote
\begin{equation}
\Delta_{s}=\sum_{p=1}^{s}\partial_{x_{p}}^{2}\;\;\;\;\mbox{for}\;\;
s=1,\ldots,n-1.
\end{equation} Then $f_j$ is determined by $f_{n-1,0}^j$ and
$f_{n-1,1}^j$ via
\begin{eqnarray}
f_{n-1,l+2}^{j}=-\Delta_{n-1}(f_{n-1,l}^{j})\;\;\;\;\mbox{for}\;\;
0\leq l\leq k-2.
\end{eqnarray} Note
\begin{eqnarray}
f_n&=&-x_n^{-1}\sum\limits_{j=1}^{n-1}x_jf_j=
-x_n^{-1}\sum\limits_{j=1}^{n-1}\sum\limits_{l=0}^kx_jf_{n-1,l}^{j}\frac{x_n^l}{l!}
\nonumber\\&=&-x_n^{-1}\sum\limits_{j=1}^{n-1}x_jf_{n-1,0}^{j}-
\sum\limits_{j=1}^{n-1}\sum\limits_{l=0}^{k-1}\frac{x_jf_{n-1,l+1}^{j}}{l+1}\frac{x_n^l}{l!}.
\end{eqnarray} Thus $f_n\in {\cal{H}}_k$ if and only if
\begin{equation} \label{eq:2}
\left\{ \begin{aligned}
         (a):\;\;\;\;\;\;\ \sum\limits_{j=1}^{n-1}x_{j}f_{n-1,0}^{j} & =0 \\
                  (b):\;\;\ \sum\limits_{j=1}^{n-1}\frac{x_jf_{n-1,l+3}^j}{l+3} & =
        -\sum\limits_{j=1}^{n-1}\Delta_{n-1}(\frac{x_{j}f_{n-1,l+1}^{j}}{l+1})
        \;\;\mbox{for}\;\;l\geq 0
                          \end{aligned} \right.
                          \end{equation} by (3.14) and (3.15). To
write down a basis of $\hat{\cal H}_{k,2}$, it is sufficient to
solve (3.16).
\begin{lemma} The following system is equivalent to (3.16).
\begin{equation} \label{eq:2}
\left\{ \begin{aligned}
         (a):\;\;\;\;\;\;\sum\limits_{j=1}^{n-1}x_{j}f_{n-1,0}^{j} & =0 \\
                  (b):\;\; \sum\limits_{j=1}^{n-1}\partial_{x_{j}}(f_{n-1,1}^{j}) & =
        -\Delta_{n-1}(\sum\limits_{j=1}^{n-1}x_{j}f_{n-1,1}^{j})
                          \end{aligned} \right.
                          \end{equation}
\end{lemma}
{\it Proof}.  Note that
\begin{eqnarray}
 &&-(l+1)\sum\limits_{j=1}^{n-1}x_{j}f_{n-1,l+3}^{j}=
 (l+3)\Delta_{n-1}(\sum\limits_{j=1}^{n-1}x_{j}f_{n-1,l+1}^{j})\nonumber\\&\Leftrightarrow&
 (l+1)\sum\limits_{j=1}^{n-1}x_{j}\Delta_{n-1}(f_{n-1,l+1}^{j})=
 (l+3)\Delta_{n-1}(\sum\limits_{j=1}^{n-1}x_{j}f_{n-1,l+1}^{j})\nonumber\\&\Leftrightarrow&
(l+1)\Delta_{n-1}(\sum\limits_{j=1}^{n-1}x_{j}f_{n-1,l+1}^{j})-
2(l+1)\sum\limits_{j=1}^{n-1}\partial_{x_{j}}(f_{n-1,l+1}^{j})=
 (l+3)\Delta_{n-1}(\sum\limits_{j=1}^{n-1}x_{j}f_{n-1,l+1}^{j})\nonumber\\&\Leftrightarrow&
(l+1)\sum\limits_{j=1}^{n-1}\partial_{x_{j}}(f_{n-1,l+1}^{j})=
-\Delta_{n-1}(\sum\limits_{j=1}^{n-1}x_{j}f_{n-1,l+1}^{j}).
\end{eqnarray} Set $l=0$ in the above equalities, we get that
\begin{eqnarray}
&&\sum\limits_{j=1}^{n-1}\frac{x_jf_{n-1,3}^j}{3}=-\sum\limits_{j=1}^{n-1}
\Delta_{n-1}(\frac{x_jf_{n-1,1}^j}{2})\nonumber\\&\Rightarrow&
\sum\limits_{j=1}^{n-1}\partial_{x_{j}}(f_{n-1,1}^{j})=
-\Delta_{n-1}(\sum\limits_{j=1}^{n-1}x_{j}f_{n-1,1}^{j}).
\end{eqnarray}
Thus (3.16) implies (3.17). But
\begin{eqnarray}
&&(l+1)\sum\limits_{j=1}^{n-1}\partial_{x_{j}}(f_{n-1,l+1}^{j})=
-\Delta_{n-1}(\sum\limits_{j=1}^{n-1}x_{j}f_{n-1,l+1}^{j})\nonumber\\&\Leftrightarrow&
-(l+1)\Delta_{n-1}(\sum\limits_{j=1}^{n-1}\partial_{x_{j}}(f_{n-1,l-1}^{j}))=
-\Delta_{n-1}(\sum\limits_{j=1}^{n-1}x_{j}\Delta_{n-1}(f_{n-1,l-1}^{j}))
\nonumber\\&\Leftrightarrow&
(l+1)\Delta_{n-1}(\sum\limits_{j=1}^{n-1}\partial_{x_{j}}(f_{n-1,l-1}^{j}))=
-\Delta_{n-1}^{2}(\sum\limits_{j=1}^{n-1}x_{j}f_{n-1,l-1}^{j})
+2\Delta_{n-1}(\sum\limits_{j=1}^{n-1}\partial_{x_{j}}(f_{n-1,l-1}^{j}))\nonumber\\&\Leftrightarrow&
(l-1)\Delta_{n-1}(\sum\limits_{j=1}^{n-1}\partial_{x_{j}}(f_{n-1,l-1}^{j}))=
-\Delta_{n-1}^{2}(\sum\limits_{j=1}^{n-1}x_{j}f_{n-1,l-1}^{j}).
\end{eqnarray}
Then by induction, one gets that (3.17(a)) implies (3.16(b)) if $l$
is odd, and (3.17(b)) implies (3.16(b)) when $l$ is even. Hence
(3.17) is equivalent to (3.16). $\qquad\Box$ \vspace{0.4cm}

Using (3.17) and the condition $f_j\in {\cal H}_k$ for
$j=1,\ldots,n-1$, we can obtain a basis of $\hat{\mathcal
{H}}_{k,2}$. For convenience, we classify these base vectors into
three disjoint subsets satisfying the following
conditions,respectively: Let $f_j$ be the first nonzero component of
the base vector
$\vec{f}=\sum\limits_{l=1}^nf_l\vs_l\in\hat{\cal{H}}_{k,2}$.

Condition (*):  $1\leq j\leq n-2$, and the powers of $x_n$ in
$f_j$ are even.

Condition (**): $1\leq j\leq n-2$, and the powers of $x_n$ in
$f_j$ are odd.

Condition (***): $j=n-1$.

To find the base vectors satisfying the condition (*), we set
$f_{n-1,1}^{l}=0$ in (3.17(b)) for $l=1,\ldots,n-1$. Moreover, for
convenience, we write
\begin{eqnarray}
f_{n-1,0}^j=\sum\limits_{l=0}^kf_{n-2,l}^j(x_1,\ldots,x_{n-2})\frac{x_{n-1}^l}{l!}
\;\;\;\;\mbox{for}\;\;j=1,\ldots,n-1.
\end{eqnarray} Continues the process. In general, we write
\begin{eqnarray}
f_{s,0}^j=\sum\limits_{l=0}^kf_{s-1,l}^j(x_1,\ldots,x_{s-1})\frac{x_{s}^l}{l!}
\;\;\;\;\mbox{for}\;\;s=1,\ldots,n-1.
\end{eqnarray}
Then by induction, the following system is equivalent to (3.17(a))
\begin{equation}\left\{\begin{array}{l}
f_{1,0}^{1}=0 \\
f_{j,0}^{j}=-x_j^{-1}\sum\limits_{r=1}^{j-1}x_{r}f_{j,0}^{r}\ \
\for\ \ j\geq2.\end{array}\right.
\end{equation}
Now we can write down those base vectors satisfying Condition (*)
by (3.23).

 From now on, the notations $r_s$ are always nonnegative integers. Assume that $f_j$ is the
first nonzero component of the base vector
$\vec{f}=\sum\limits_{l=1}^nf_l\vs_l\in\hat{\cal{H}}_{k,2}$. Then
$f_{j,0}^j=0$ by (3.23). We take $f_{n-1,0}^j$ to be the following
monomials:
\begin{eqnarray}
f_{n-1,0}^j=x_1^{r_1}\cdots x_{n-1}^{r_{n-1}},
\end{eqnarray}
where $r_1+\cdots+r_{n-1}=k$ and $r_{j+1}+\cdots+r_{n-1}>0.$
Moreover, we set
\begin{eqnarray}
f_{n-1,0}^q=f_{q,0}^q\;\;\;\;\for\;\;j<q\leq n-1.
\end{eqnarray}
Thus
\begin{eqnarray}
f_{n-1,0}^q=-\delta_{r_1+\cdots+r_q,k}(\sum\limits_{s=1}^{n-1}\delta_{r_{q},s})\frac{x_{j}}{x_{q}}
\prod\limits_{s=1}^{n-1} x_{s}^{r_{s}},
\end{eqnarray} and so
\begin{eqnarray}
f_{q}=\sum\limits_{l=0}^{\llbracket\frac{k}{2}\rrbracket}(-1)^{l+1}\delta_{r_1+\cdots+r_q,k}
(\sum\limits_{s=1}^{n-1}\delta_{r_{q},s})\Delta_{n-1}^{l}\left(\frac{x_{j}x_{n}^{2l}}{(2l)!x_{q}}
\prod\limits_{s=1}^{n-1} x_{s}^{r_{s}}\right)
\end{eqnarray}
for $j< q\leq n-1.$ Then we obtain that: \vspace{0.4cm}

 (I). The following vectors are the
base vectors satisfying  Condition (*):
\begin{equation}
\vec{f}=\sum\limits_{l=j}^{n-1}f_l(\vs_l-x_n^{-1}x_l\vs_n)
\end{equation}
for some $j\in \{1,...,n-2\}$, where
\begin{equation}
f_{j}=\sum\limits_{l=0}^{\llbracket\frac{k}{2}\rrbracket}(-1)^{l}\Delta_{n-1}^{l}\left(\frac{x_{n}^{2l}}{(2l)!}
\prod\limits_{s=1}^{n-1} x_{s}^{r_{s}}\right)\end{equation} and
$f_{j+1},...,f_{n-1}$ are given in (3.27) for any nonnegative
integers $r_1,...,r_{n-1}$ with $r_1+\cdots+r_{n-1}=k$ and
$r_{j+1}+\cdots+r_{n-1}>0.$ \vspace{0.4cm}

To get the base vectors satisfying condition (**) and (***), we set
$f_{n-1,0}^l=0$ in (3.17(a)), and simplify (3.17(b)). Note that it
can be written as
\begin{eqnarray}
\partial_{x_{n-1}}(f_{n-1,1}^{n-1})+\Delta_{n-1}(x_{n-1}f_{n-1,1}^{n-1})=
-\sum\limits_{l=1}^{n-2}\partial_{x_{l}}(f_{n-1,1}^{l})
-\Delta_{n-1}(\sum\limits_{l=1}^{n-2}x_{l}f_{n-1,1}^{l})
\end{eqnarray}
We write
\begin{eqnarray}
f_{n-1,1}^{s}=\sum\limits_{l=0}^{k-1}g_{l}^{s}(x_{1},\ldots,x_{n-2})\frac{x_{n-1}^{l}}{l!}
\;\;\;\;\for\;\; 1\leq s\leq n-1.
\end{eqnarray}
Substituting (3.31) to (3.30), we get
\begin{eqnarray}
\mbox{left}&=&
\partial_{x_{n-1}}(\sum\limits_{l=0}^{k-1}g_{l}^{n-1}\frac{x_{n-1}^{l}}{l!})+
(\Delta_{n-2}+\partial_{x_{n-1}}^{2})(x_{n-1}
\sum\limits_{l=0}^{k-1}g_{l}^{n-1}\frac{x_{n-1}^{l}}{l!})\nonumber\\&=&
\sum\limits_{l=0}^{k-2}(l+3)g_{l+1}^{n-1}\frac{x_{n-1}^{l}}{l!}+
\sum\limits_{l=0}^{k}l\Delta_{n-2}(g_{l-1}^{n-1}\frac{x_{n-1}^{l}}{l!})
\end{eqnarray} and
\begin{eqnarray}
\mbox{right}&=&-\sum\limits_{s=1}^{n-2}\partial_{x_{s}}
(\sum\limits_{l=0}^{k-1}g_{l}^{s}\frac{x_{n-1}^{l}}{l!})-
(\Delta_{n-2}+\partial_{x_{n-1}}^{2})
(\sum\limits_{s=1}^{n-2}\sum\limits_{l=0}^{k-1}x_{s}g_{l}^{s}\frac{x_{n-1}^{l}}{l!})
\nonumber\\&=&
-\sum\limits_{l=0}^{k-1}\sum\limits_{s=1}^{n-2}(\Delta_{n-2}x_{s}+\partial_{x_{s}})
(g_{l}^{s}\frac{x_{n-1}^{l}}{l!})-\sum\limits_{l=0}^{k-3}
\sum\limits_{s=1}^{n-2}x_{s}g_{l+2}^{s}\frac{x_{n-1}^{l}}{l!}.
\end{eqnarray}
Thus we obtain
\begin{eqnarray}
g_{l+1}^{n-1}=-\frac{l}{l+3}\Delta_{n-2}(g_{l-1}^{n-1})-
\frac{1}{l+3}\sum\limits_{s=1}^{n-2}(\Delta_{n-2}x_{s}+\partial_{x_{s}})(g_{l}^{s})-
\frac{1}{l+3}\sum\limits_{s=1}^{n-2}x_{s}g_{l+2}^{s}
\end{eqnarray}
for $0\leq l\leq k-2$, where $g_{-1}^{s}=g_{k}^{s}=0$. If we define
$(-1)!!=0$, then
\begin{eqnarray}\nonumber
g_{2l}^{n-1}&=&
\sum\limits_{r=1}^{n-2}\sum\limits_{s=1}^{l}(-1)^{l-s+1}
\frac{(2s-2)!!(2l-1)!!}{(2s-1)!!(2l+2)!!}(\Delta_{n-2}^{l-s+1}x_{r}+
2s\Delta_{n-2}^{l-s}\partial_{x_{r}})(g_{2s-1}^{r})\\
&&+(-1)^{l}\frac{(2l-1)!!}{(2l+2)!!}\Delta_{n-2}^{l}
(2g_{0}^{n-1}+\sum\limits_{r=1}^{n-2}x_{r}g_{1}^{r})
-\frac{1}{2l+2}\sum\limits_{r=1}^{n-2}x_{r}g_{2l+1}^{r}
\end{eqnarray} and
\begin{eqnarray}
\nonumber g_{2l+1}^{n-1}&=&
\sum\limits_{r=1}^{n-2}\sum\limits_{s=0}^{l}(-1)^{l-s}
\frac{(2s-1)!!(2l)!!}{(2s)!!(2l+3)!!}(\Delta_{n-2}^{l-s+1}x_{r}+
(2s+1)\Delta_{n-2}^{l-s}\partial_{x_{r}})(g_{2s}^{r})\\&&
-\frac{1}{2l+3}\sum\limits_{r=1}^{n-2}x_{r}g_{2l+2}^{r}
\end{eqnarray}
for $l\geq 0$. The above two equalities tell us that $f_{n-1}$ is
determined by $f_r$ whenever $r\leq n-2$ and $g_0^{n-1}$ under the
condition $f_{n-1,0}^l=0$ for $l=1,\ldots,n-1$. Moreover, $f_r$ is
determined by $g_s^r(x_1,\ldots,x_{n-2})$ which can be any
homogenous polynomial with degree $k-1-s$. These help us to write
down those base vectors satisfying condition (**) and (***).

Setting $g_0^{n-1}=0$, similarly as (I), we get that \vspace{0.4cm}

(II). The following vectors are the base vectors satisfying
Condition (**):
\begin{eqnarray}\vec
f=f_j\vs_j+f_{n-1}\vs_{n-1}-x_n^{-1}(x_jf_j+x_{n-1}f_{n-1})\vs_n
\end{eqnarray}
for some $j\in \{1,...,n-2\}$, where
\begin{eqnarray}
f_{j}=
\sum\limits_{l=0}^{\llbracket\frac{k}{2}\rrbracket}(-1)^{l}\Delta_{n-1}^{l}\left(\frac{x_{n}^{2l+1}}{(2l+1)!}\prod\limits_{s=1}^{n-1}
x_{s}^{r_{s}}\right)
\end{eqnarray}
\begin{eqnarray}& &
f_{n-1}=\sum\limits_{l=0}^{\llbracket\frac{k}{2}\rrbracket}(-1)^{l+1}
\{\sum\limits_{p=1}^{k-1}\frac{\delta_{r_{n-1},p}p!}{(p+1)(p-1)!(2l+1)!}
\Delta_{n-1}^{l}\left(x_{j}x_{n-1}^{p-1}x_{n}^{2l+1}\prod\limits_{s=1}^{n-2}
x_{s}^{r_{s}}\right)\nonumber\\\nonumber&&
+\sum\limits_{q=0}^{\llbracket\frac{k}{2}\rrbracket}[
\frac{\delta_{r_{n-1},1}(-1)^{q+1}(2q-1)!!}{(2q+2)!!(2q)!(2l+1)!}\Delta_{n-1}^{l}\Delta_{n-2}^{q}
\left(x_{j}x_{n-1}^{2q+1}x_{n}^{2l+1}\prod\limits_{s=1}^{n-2}x_{s}^{r_{s}}\right)\\\nonumber&&
+\frac{(-1)^{q+1}(r_{n-1}-1)!!(r_{n-1}+2q)!!r_{n-1}!}{r_{n-1}!!(r_{n-1}+2q+3)!!(r_{n-1}+2q+1)!(2l+1)!}
(\Delta_{n-1}^{l}\Delta_{n-2}^{q+1}
\left(x_{j}x_{n-1}^{2q+1}x_{n}^{2l+1}\prod\limits_{s=1}^{n-1}x_{s}^{r_{s}}\right)\\&&
+(r_{n-1}+1)r_{j}\Delta_{n-1}^{l}\Delta_{n-2}^{q}
\left(x_{j}^{-1}x_{n-1}^{2q+1}x_{n}^{2l+1}\prod\limits_{s=1}^{n-1}x_{s}^{r_{s}}\right))]\}
\end{eqnarray}
for any nonnegative integers $r_1,...,r_{n-1}$ with
$r_1+\cdots+r_{n-1}=k-1$. \vspace{0.4cm}

We set $f_r=0$ for $r\leq n-2$ and
\begin{equation}
g_0^{n-1}=x_1^{r_1}\cdots x_{n-2}^{r_{n-2}},
\end{equation} where
$r_1+\cdots+r_{n-2}=k-1.$ Then we get \vspace{0.4cm}

(III). The following vectors are the base vectors satisfying
Condition (***):
\begin{eqnarray}\vec f=f_{n-1}(\vs_{n-1}-x_n^{-1}x_{n-1}\vs_{n-1}),
\end{eqnarray}
where
\begin{eqnarray}
f_{n-1}=\sum\limits_{l=0}^{\llbracket\frac{k}{2}\rrbracket}\sum\limits_{q=0}^{\llbracket\frac{k}{2}\rrbracket}\frac{(-1)^{l+q}
2(2q-1)!!}{(2q+2)!!(2q)!(2l+1)!}\Delta_{n-1}^{l}\Delta_{n-2}^{q}
\left(x_{n-1}^{2q}x_{n}^{2l+1}\prod\limits_{s=1}^{n-2}x_{s}^{r_{s}}\right)
\end{eqnarray}
for any nonnegative integers $r_1,...,r_{n-2}$ with
$r_1+\cdots+r_{n-2}=k-1$.

\begin{proposition} The set of the vectors $\vec f$ given in
(3.27)-(3.29), (3.37)-(3.39) and (3.41)-(3.42) forms a basis of
$\hat{\cal H}_{k,2}$.
\end{proposition}

Now we give bases of $\hat{\cal H}_{k,2\mp}$ in the case of $n=4$.
Recall that the representation operators of negative simple root
vectors of $o'(4,\mathbb{C})$ are
\begin{equation}
f_{\alpha_1}=y_{2}\partial_{y_{1}}-y_{3}\partial_{y_{4}}+
\kappa_{2}\partial_{\kappa_{1}}-\kappa_{3}\partial_{\kappa_{4}}
\end{equation} and
\begin{equation}
f_{\alpha_2}=y_{4}\partial_{y_{1}}-y_{3}\partial_{y_{2}}+
\kappa_{4}\partial_{\kappa_{1}}-\kappa_{3}\partial_{\kappa_{2}}.
\end{equation}
Then
\begin{equation}
\hat{\cal H}_{k,2-}=\mbox{span}\
\{\mbox{Re}(f_{\alpha_2}^sf_{\alpha_1}^r(v_{2,-})),\
\mbox{Im}(f_{\alpha_2}^sf_{\alpha_1}^r(v_{2,-}))\mid\ r,s\
\in\mathbb{N}\}
\end{equation} by Lemma 2.4. Observe that
\begin{eqnarray}\nonumber
f_{\alpha_1}^{r}(v_{2,-})&=&(y_{2}\partial_{y_{1}}+\kappa_{2}\partial_{\kappa_{1}})^{r}
(-y_{2}y_{1}^{k-1}\kappa_{1}+y_{1}^{k}\kappa_{2})\\\nonumber
&=&(y_{2}^{r}\partial_{y_{1}}^{r}+ry_{2}^{r-1}\kappa_{2}\partial_{y_{1}}^{r-1}\partial_{\kappa_{1}})
(-y_{2}y_{1}^{k-1}\kappa_{1}+y_{1}^{k}\kappa_{2})\\&=&
(\prod\limits_{j=0}^{r-1}k-1-j)(-y_{2}^{r+1}y_{1}^{k-1-r}\kappa_{1}+y_{2}^{r}y_{1}^{k-r}\kappa_{2})
\end{eqnarray}
for $r\leq k-1$, and
\begin{eqnarray}
&&f_{\alpha_2}^{s}(-y_{2}^{r+1}y_{1}^{k-1-r}\kappa_{1}+y_{2}^{r}y_{1}^{k-r}\kappa_{2})\nonumber\\&=&
(y_{4}\partial_{y_{1}}-y_{3}\partial_{y_{2}}+
\kappa_{4}\partial_{\kappa_{1}}-\kappa_{3}\partial_{\kappa_{2}})^{s}
(-y_{2}^{r+1}y_{1}^{k-1-r}\kappa_{1}+y_{2}^{r}y_{1}^{k-r}\kappa_{2})\nonumber\\&=&
((y_{4}\partial_{y_{1}}-y_{3}\partial_{y_{2}})^{s}+
s(y_{4}\partial_{y_{1}}-y_{3}\partial_{y_{2}})^{s-1}
(\kappa_{4}\partial_{\kappa_{1}}-\kappa_{3}\partial_{\kappa_{2}}))
(-y_{2}^{r+1}y_{1}^{k-1-r}\kappa_{1}+y_{2}^{r}y_{1}^{k-r}\kappa_{2})\nonumber\\&=&
\frac{1}{2}\left(\begin{array}{l}
-(y_{4}\partial_{y_{1}}-y_{3}\partial_{y_{2}})^{s}y_{2}^{r+1}y_{1}^{k-1-r}-
is(y_{4}\partial_{y_{1}}-y_{3}\partial_{y_{2}})^{s-1}y_{2}^{r}y_{1}^{k-r}\\
(y_{4}\partial_{y_{1}}-y_{3}\partial_{y_{2}})^{s}y_{2}^{r}y_{1}^{k-r}-
is(y_{4}\partial_{y_{1}}-y_{3}\partial_{y_{2}})^{s-1}y_{2}^{r+1}y_{1}^{k-1-r}\\
-i(y_{4}\partial_{y_{1}}-y_{3}\partial_{y_{2}})^{s}y_{2}^{r+1}y_{1}^{k-1-r}-
s(y_{4}\partial_{y_{1}}-y_{3}\partial_{y_{2}})^{s-1}y_{2}^{r}y_{1}^{k-r}\\
i(y_{4}\partial_{y_{1}}-y_{3}\partial_{y_{2}})^{s}y_{2}^{r}y_{1}^{k-r}-
s(y_{4}\partial_{y_{1}}-y_{3}\partial_{y_{2}})^{s-1}y_{2}^{r+1}y_{1}^{k-1-r}
\end{array}\right)
\end{eqnarray}
for $0\leq s\leq k+1$. Denote
\begin{equation}
g_+(r,s)=2^{k}\mbox{Re}((y_{4}\partial_{y_{1}}-y_{3}\partial_{y_{2}})^{s}y_{2}^{r}y_{1}^{k-r})
\end{equation}
and
\begin{equation}
g_-(r,s)=2^{k}\mbox{Im}((y_{4}\partial_{y_{1}}-y_{3}\partial_{y_{2}})^{s}y_{2}^{r}y_{1}^{k-r}).
\end{equation} Then the real part of (3.47) is
\begin{eqnarray}
\vec{v}(r,s)=\frac{1}{2^{k+1}}\left(\begin{array}{c}
        -g_+(r+1,s)+s g_-(r,s-1)\\
        g_+(r,s)+s g_-(r+1,s-1)\\
        g_-(r+1,s)-s g_+(r,s-1)\\
        -g_-(r,s)-s g_+(r+1,s-1)
        \end{array}\right),
        \end{eqnarray} and the imaginary part of (3.47) is
 \begin{eqnarray}
\vec{w}(r,s)=\frac{1}{2^{k+1}}\left(\begin{array}{c}
        -g_-(r+1,s)-s g_+(r,s-1)\\
        g_-(r,s)-s g_+(r+1,s-1)\\
        -g_+(r+1,s)-s g_-(r,s-1)\\
        g_+(r,s)-s g_-(r+1,s-1)
        \end{array}\right).
\end{eqnarray}
Moreover, a straightforward calculation shows that

i) If $0\leq r+s\leq k$ and $0\leq r< s$, then
\begin{eqnarray}
g_{\pm}(r,s)&=&\sum\limits_{l=0}^{s}\sum\limits_{p=0}^{k-r-s}\sum\limits_{q=0}^{s-r}
\delta_{(-1)^{r+p+q},\pm1}(-1)^{l+\llbracket\frac{r+p+q}{2}\rrbracket}s!
{k-r\choose s-l}{r\choose l}{k-r-s\choose p}{s-r\choose q}
\nonumber\\&&\times
(x_{1}^{2}+x_{3}^{2})^{l}(x_{2}^{2}+x_{4}^{2})^{r-l}
x_{1}^{k-r-s-p}x_{2}^{q}x_{3}^{p}x_{4}^{s-r-q}.
\end{eqnarray}

ii) If $0\leq r+s\leq k$ and $0\leq s\leq r$, then
\begin{eqnarray}
g_{\pm}(r,s)&=&\sum\limits_{l=0}^{s}\sum\limits_{p=0}^{k-r-s}\sum\limits_{q=0}^{r-s}
\delta_{(-1)^{r+p+q},\pm1}(-1)^{l+\llbracket\frac{r+p+q}{2}\rrbracket}s!
{k-r\choose s-l}{r\choose l}{k-r-s\choose p}{r-s\choose
q}\nonumber\\&&\times
(x_{1}^{2}+x_{3}^{2})^{l}(x_{2}^{2}+x_{4}^{2})^{s-l}
x_{1}^{k-r-s-p}x_{2}^{r-s-q}x_{3}^{p}x_{4}^{q}.
\end{eqnarray}

iii) If $r+s>k$ and $0\leq r<s$, then
\begin{eqnarray}
g_{\pm}(r,s)&=&\sum\limits_{l=0}^{s}\sum\limits_{p=0}^{r+s-k}\sum\limits_{q=0}^{s-r}
\delta_{(-1)^{r-p+q},\pm1}(-1)^{l+\llbracket\frac{r-p+q}{2}\rrbracket}s!
{k-r\choose s-l}{r\choose l}{r+s-k\choose p}{s-r\choose q}
\nonumber\\&&\times
(x_{1}^{2}+x_{3}^{2})^{k-r-s+l}(x_{2}^{2}+x_{4}^{2})^{r-l}
x_{1}^{r+s-k-p}x_{2}^{q}x_{3}^{p}x_{4}^{s-r-q}.
\end{eqnarray}

iv) If $r+s>k$ and $0\leq s\leq r$, then
\begin{eqnarray}
g_{\pm}(r,s)&=&\sum\limits_{l=0}^{s}\sum\limits_{p=0}^{r+s-k}\sum\limits_{q=0}^{r-s}
\delta_{(-1)^{r-p-q},\pm1}(-1)^{l+\llbracket\frac{r-p-q}{2}\rrbracket}s!
{k-r\choose s-l}{r\choose l}{r+s-k\choose p}{r-s\choose
q}\nonumber\\&&\times
(x_{1}^{2}+x_{3}^{2})^{k-r-s+l}(x_{2}^{2}+x_{4}^{2})^{s-l}
x_{1}^{r+s-k-p}x_{2}^{q}x_{3}^{p}x_{4}^{r-s-q}.
\end{eqnarray}

Now we have the following base vectors of $\hat{\mathcal
{H}}_{k,2-}$:

i) \begin{eqnarray} \vec{f}=\vec{v}(r,s),\;\;\;\;   \mbox{if}\;
\left\{\begin{array}{l}
 0\leq r\leq\llbracket\frac{k}{2}\rrbracket\\
 0\leq s\leq\llbracket\frac{k+1}{2}\rrbracket
 \end{array}\right.\;\;\;\;  \mbox{or}\;
 \left\{\begin{array}{l}
 \llbracket\frac{k}{2}\rrbracket+1\leq r\leq k-1\\
 0\leq s\leq\llbracket\frac{k}{2}\rrbracket
 \end{array}\right.,
 \end{eqnarray}

 ii)\begin{eqnarray}
 \vec{f}=\vec{w}(r,s),\;\;\;\;   \mbox{if}\; \left\{\begin{array}{l}
 0\leq r\leq\llbracket\frac{k}{2}\rrbracket\\
 0\leq s\leq\llbracket\frac{k+1}{2}\rrbracket
 \end{array}\right.\;\;\;\;  \mbox{or}\;
 \left\{\begin{array}{l}
 \llbracket\frac{k}{2}\rrbracket+1\leq r\leq k-1\\
 0\leq s\leq\llbracket\frac{k}{2}\rrbracket
 \end{array}\right.,
 \end{eqnarray}

iii) \begin{eqnarray}
 \vec{f}=\left\{\begin{array}{ll}
 \vec{v}(r,s)& \mbox{if}\  \vec{v}(r,s)\neq 0\\
 \vec{w}(r,s)& \mbox{if}\  \vec{v}(r,s)= 0
 \end{array}\right.\;\; \mbox{for}\;
 r=\frac{k-1}{2},\; s=\frac{k+1}{2}\; \mbox{and}\ k\ \mbox{ is odd}.
\end{eqnarray}
Then by (3.45), we get that
\begin{proposition} The set of vectors $\vec f$ given in (3.56)-(3.58) forms
a basis of $\hat{\mathcal {H}}_{k,2-}$.
\end{proposition}

\begin{remark} We observe that there  exists a linear isomorphism between vector
space $\hat{\cal H}_{k,2-}$ and $\hat{\cal H}_{k,2+}$
$$
\sigma: \hat{\cal H}_{k,2-}\longrightarrow \hat{\cal
H}_{k,2+},\;\;\mbox{where}\;\; \sigma(x_{1})= x_{1},\;
\sigma(x_{2})= x_{4},\; \sigma(x_{3})= x_{3}\;\mbox{and}\;
\sigma(x_{4})= x_{2}.
$$
These give a basis of $\hat{\cal H}_{k,2+}$  by easily interchanging
$x_{2}$ and $x_{4}$ in a basis of $\hat{\cal H}_{k,2-}$.
\end{remark}

At the end of this section, we will use Xu's method in [X1] to
construct a uniform basis of the polynomial solution space of Navier
equations, whose the cardinality was pre-used in the proof of
Theorem 2.12. It is different from those bases given above. It is
not listed in accordance to the irreducible summands of the
polynomial solution space. Xu's method is also critical to solve the
initial value problems of Navier equations and Lam\'{e} equations in
the next section.

\begin{lemma}{[X1]}
Suppose that $\mathscr{A}$ is a free module of a subalgebra
$\mathscr{B}$ generated by a filtrated subspace
$V=\bigcup\limits_{r=0}^{\infty}V_{r}$ (i.e. $V_{r}\subset
V_{r+1}$). Let $T_{0}$ be a linear operator on $\mathscr{A}$ with
right inverse $T_{0}^{-}$ such that $T_{0}(\mathscr{B})\subset
\mathscr{B}$, $T_{0}^{-}(\mathscr{B})\subset \mathscr{B}$ and
$T_{0}(\eta_{1}\eta_{2})=T_{0}(\eta_{1})\eta_{2}$ for
$\eta_{1}\in\mathscr{B}$, $\eta_{2}\in V$. Let $T_{1},\ldots,T_{m}$
be linear operators on $\mathscr{A}$ such that $T_{j}(V)\subset V$,
$T_{j}(f\zeta)=fT_{j}(\zeta)$ for $j=1,\ldots,m$, $f\in
\mathscr{B}$, $\zeta\in\mathscr{A}$. If $T_{0}^{m}(h)=0$ with
$h\in\mathscr{B}$ and $g\in V$, then
\begin{equation}
u=\sum\limits_{j=0}^{\infty}(\sum\limits_{s=1}^{m}(T_{0}^{-})^{s}T_{s})^{j}(hg)
\end{equation}
is a solution of the equation
\begin{equation}
(T_{0}^{m}-\sum\limits_{j=1}^{m}T_{0}^{m-j}T_{j})(u)=0
\end{equation}
Suppose $T_{j}(V_{r})\subset V_{r-1}$ for $j=1,\cdots,m$,
$r\in\mathbb{N}$, where $V_{-1}=0$. Then any polynomial solution of
(3.60) is a linear combinations of the solutions of the form (3.59).
In particular, if $T_{r}T_{s}=T_{s}T_{r}$, $T_{0}T_{j}=T_{j}T_{0}$
and $T_{0}^{-}T_{j}=T_{j}T_{0}^{-}$ for any
$j,r,s\in\{1,\ldots,m\}$, then $u$ can be written as follows:
\begin{eqnarray}
u=\sum\limits_{i_{1},\ldots,i_{m}=0}^{\infty}{i_{1}+\cdots+i_{m}\choose
i_{1},\cdots,i_{m}}
(T_{0}^{-})^{\sum\limits_{s=1}^{m}si_{s}}(h)(\prod\limits_{r=1}^{m}T_{r}^{i_{r}})(g).
\end{eqnarray}
\end{lemma}

Note that Navier equations (1.6) can be written as the following
form
\begin{equation}
(T_{0}^{2}-T_{0}T_{1}-T_{2})(\vec u)=0,
\end{equation}
where
\begin{equation}
T_{0}=\partial_{x_{1}}I_{n},\ \ \ \ \ \ \ \ T_{1}=-\left(
\begin{array}{cccc}
0&\frac{b}{b+1}\partial_{x_{2}}&\cdots&\frac{b}{b+1}\partial_{x_{n}}\\
b\partial_{x_{2}}&0&\cdots&0\\
\vdots&\vdots&\ddots&\vdots\\
b\partial_{x_{n}}&0&\cdots&0\end{array} \right),
\end{equation}
\begin{eqnarray}
T_{2}=-\left(
\begin{array}{ccccc}
\frac{1}{b+1}\sum\limits_{j=2}^{n}\partial_{x_{j}}^{2}&0& 0&\cdots &0\\
0&b\partial_{x_{2}}^2+\sum\limits_{j=2}^{n}\partial_{x_{j}}^{2}
&b\partial_{x_{2}}\partial_{x_{3}}&\cdots &b\partial_{x_{2}}\partial_{x_{n}}\\
0&b\partial_{x_{3}}\partial_{x_{2}}
&b\partial_{x_{3}}^2+\sum\limits_{j=2}^{n}\partial_{x_{j}}^{2}
&\ldots
&b\partial_{x_{3}}\partial_{x_{n}}\\
\vdots&\vdots&\vdots&\ddots&\vdots\\
0&b\partial_{x_{n}}\partial_{x_{2}}
&b\partial_{x_{n}}\partial_{x_{3}}&\cdots
&b\partial_{x_{n}}^2+\sum\limits_{j=2}^{n}\partial_{x_{j}}^{2}
\end{array}
\right),
\end{eqnarray}
where $b=(\iota_1+\iota_2)/\iota_1$. Set
$\mathscr{B}=\mbb{R}[x_{1}]I_n$ and
\begin{eqnarray}
V=\sum_{r=1}^n\mbb{R}[x_{2},\ldots,x_{n}]\vs_r.
\end{eqnarray} By Lemma 3.7,  any polynomial solution of
(3.62) is  a linear combination of
\begin{equation}
\vec
u=\sum\limits_{m=0}^{\infty}((T_{0}^{-})^{m}(T_{1}+T_{0}^{-}T_{2})^{m})(hg),
\end{equation}
where $h=x_{1}^{\epsilon}I_n$ and $g\in V$. In order to write down
these solutions explicitly, we need to calculate
$(T_{0}^{-})^{m}(T_{1}+T_{0}^{-}T_{2})^{m}.$ Recall that $E_{r,s}$
is the $n\times n$ matrix whose $(r,s)$-th entry is $1$ and the
others are $0$. We define the linear operator $\int_{(x_{1})}$ on
$\mbb{R}[x_1,\ldots,x_{n}]$ by
\begin{equation}\int_{(x_{1})}(x_1^{r_1}x_2^{r_2}\cdots
x_n^{r_n})=\frac{1}{r_1+1}x_1^{r_1+1}x_2^{r_2}\cdots
x_n^{r_n}.\end{equation} We take
\begin{equation}T_0^-=\int_{(x_{1})}I_n.\end{equation}
Then
\begin{eqnarray}
&&-(T_1+T_0^-T_2)\nonumber\\
&=&((b+1)^{-1}\sum\limits_{j=2}^{n}\partial_{x_{j}}^{2}\int_{(x_{1})})E_{1,1}+
\sum\limits_{r=2}^n(\frac{b\partial_{x_r}}{b+1}E_{1,r}+b\partial_{x_r}E_{r,1})\nonumber\\&&
+\sum\limits_{r=2}^n(\sum\limits_{j=2}^{n}\partial_{x_{j}}^{2}\int_{(x_{1})})E_{r,r}+
\sum\limits_{r,s=2}^n(b\partial_{x_r}\partial_{x_s}\int_{(x_{1})})E_{r,s}.
\end{eqnarray}
Observe that $\int_{(x_{1})}$,
$\partial_{x_2},\ldots,\partial_{x_{n}}$
 commute pairwise, and so the entries of
$(T_{0}^{-})^{m}(T_{1}+T_{0}^{-}T_{2})^{m}$ are polynomials in
$\mbb{R}[\int_{(x_{1})},\partial_{x_2},\ldots,\partial_{x_{n}}]$. In
order to use linear algebra, we replace these operators by real
numbers as Xu did in [X2]. Indeed, if we let
\begin{eqnarray}
B(\hat a_1,\hat a_2,...,\hat a_n)=\left(
\begin{array}{ccccc}
\frac{\xi}{b+1}&
\frac{b\hat a_{2}}{b+1}&\frac{b\hat a_{3}}{b+1}&\cdots&\frac{b\hat a_{n}}{b+1}\\
b\hat a_{2}&b\hat a_{1}\hat a_{2}^{2}+\xi&b\hat a_{1}\hat a_{2}\hat a_{3}&\cdots&b\hat a_{1}\hat a_{2}\hat a_{n}\\
b\hat a_{3}&b\hat a_{1}\hat a_{3}\hat a_{2}&b\hat a_{1}\hat a_{3}^{2}+\xi&\cdots&b\hat a_{1}\hat a_{3}\hat a_{n}\\
\vdots&\vdots&\vdots&\ddots&\vdots\\
b\hat a_{n}&b\hat a_{1}\hat a_{n}\hat a_{2}&b\hat a_{1}\hat
a_{n}\hat a_{3}&\cdots&b\hat a_{1}\hat a_{n}^{2}+\xi
\end{array}
\right)
\end{eqnarray}
with $\hat a_s\in\mbb{R}$,
\begin{equation}\eta=\sum\limits_{j=2}^{n}\hat
a_{j}^{2}\qquad\mbox{and}\qquad \xi=\hat a_{1}\eta,\end{equation}
then
\begin{equation}B\left(\int_{(x_{1})},\partial_{2},\ldots,\partial_{x_{n}}\right)
=-(T_1+T_0^-T_2).\end{equation}

Observe that for $m\geq1$,
\begin{eqnarray}
B^{m}=\left(
\begin{array}{cc}
\frac{1}{\sqrt{b+1}}&0\\0&I_{n-1}
\end{array} \right)A^{m}\left(
\begin{array}{cc}
\sqrt{b+1}&0\\0&I_{n-1}
\end{array}
\right)
\end{eqnarray}
 with
\begin{eqnarray}
A=\left(
\begin{array}{ccccc}
\frac{\xi}{b+1}&
\frac{b\hat a_{2}}{\sqrt{b+1}}&\frac{b\hat a_{3}}{\sqrt{b+1}}&\cdots&\frac{b\hat a_{n}}{\sqrt{b+1}}\\
\frac{b\hat a_{2}}{\sqrt{b+1}}&b\hat a_{1}\hat a_{2}^{2}+\xi&b\hat a_{1}\hat a_{2}\hat a_{3}&\cdots&b\hat a_{1}\hat a_{2}\hat a_{n}\\
\frac{b\hat a_{3}}{\sqrt{b+1}}&b\hat a_{1}\hat a_{3}\hat a_{2}&b\hat a_{1}\hat a_{3}^{2}+\xi&\cdots&b\hat a_{1}\hat a_{3}\hat a_{n}\\
\vdots&\vdots&\vdots&\ddots&\vdots\\
\frac{b\hat a_{n}}{\sqrt{b+1}}&b\hat a_{1}\hat a_{n}\hat a_{2}&b\hat
a_{1}\hat a_{n}\hat a_{3}&\cdots&b\hat a_{1}\hat a_{n}^{2}+\xi
\end{array}
\right)
\end{eqnarray} is symmetrical (note that
$b+1=(2\iota_1+\iota_2)/\iota_1>0$). Moreover, we denote
\begin{equation}a=(b+1)^{2}+1\qquad\mbox{and}\qquad
\varpi=(b+2)^{2}\xi^{2}+4(b+1)\eta.\end{equation} It can be proved
that the eigenvalues of $A$ are
\begin{eqnarray}
\xi,\;\; \theta_1=\frac{a\xi+b\sqrt{\varpi}}{2(b+1)}\;\;
\mbox{and}\;\; \theta_2=\frac{a\xi-b\sqrt{\varpi}}{2(b+1)},
\end{eqnarray}
where the multiplicity of $\xi$ is $n-2$. Recall that
$\vs_r=(0,...,0,\stl{r}{1},0,...,0)^T$. We can take  orthonormal
eigenvectors $\vec{v}_{j}=(0,p_{j,2},\ldots,p_{j,n})^T$ for
$j=1,...,n-2$ corresponding to the eigenvalue $\xi$, a unit
eigenvector
\begin{eqnarray}
\vec{v}_{n-1}=\frac{\sqrt{2(b+1)\eta}}{\sqrt{\varpi+(b+2)\xi\sqrt{\varpi}}}\vs_1+
\sum\limits_{r=2}^n\frac{\hat a_{r}((b+2)\xi+\sqrt{\varpi})}
{\sqrt{2\eta(\varpi+(b+2)\xi\sqrt{\varpi})}}\vs_r
\end{eqnarray} corresponding to the eigenvalue $\theta_1$,
 and  a unit eigenvector
\begin{eqnarray}
\vec{v}_{n}=\frac{\sqrt{2(b+1)\eta}}{\sqrt{\varpi-(b+2)\xi\sqrt{\varpi}}}\vs_1+
\sum\limits_{r=2}^n\frac{\hat a_{r}((b+2)\xi-\sqrt{\varpi})}
{\sqrt{2\eta(\varpi-(b+2)\xi\sqrt{\varpi})}}\vs_r.
\end{eqnarray}
corresponding to the eigenvalue $\theta_2$. Setting
$P=(\vec{v}_{1},\cdots,\vec{v}_{n})$ and
\begin{eqnarray}
J=\left(
\begin{array}{ccc}
\xi I_{n-2}&0&0\\0&\theta_{1}&0\\0&0&\theta_{2}
\end{array}
\right),
 \end{eqnarray}  we have that
\begin{eqnarray}
A^{m}=PJ^{m}P^{t}=P\cdot\mbox{diag}(\xi^{m},\cdots,\xi^{m},\theta_{1}^{m},\theta_{2}^{m})
\cdot P^{T}=(c_{r,s})_{n\times n},
\end{eqnarray}
where
\begin{equation}
c_{1,j}=c_{j,1}=p_{1,n-1}p_{j,n-1}\theta_{1}^{m}+p_{1,n}p_{j,n}\theta_{2}^{m}
\end{equation} for $1\leq j\leq n$, and
\begin{eqnarray}\nonumber
c_{r,s}&=&\sum\limits_{j=1}^{n-2}p_{r,j}p_{s,j}\xi^{m}+
p_{r,n-1}p_{s,n-1}\theta_{1}^{m}+p_{r,n}p_{s,n}\theta_{2}^{m}\\
&=& \delta_{r,s}\xi^{m}+p_{r,n-1}p_{s,n-1}(\theta_{1}^{m}-\xi^{m})
+p_{r,n}p_{s,n}(\theta_{2}^{m}-\xi^{m}),
\end{eqnarray}
for $r,s\in \{2,\ldots,n\}$ by the the fact $PP^T=I_n$. Substituting
(3.76) into the above two equations, we have that
\begin{eqnarray}
\nonumber c_{11}&=&\frac{2(b+1)\eta}{\varpi+(b+2)\xi
\sqrt{\varpi}}\cdot\frac{(a\xi+b\sqrt{\varpi})^{m}}{2^{m}(b+1)^{m}}+
\frac{2(b+1)\eta}{\varpi-(b+2)\xi
\sqrt{\varpi}}\cdot\frac{(a\xi-b\sqrt{\varpi})^{m}}{2^{m}(b+1)^{m}}\\\nonumber&=&
\frac{1}{2^{m}(b+1)^{m}}(\sum\limits_{2\mid j} {m\choose
j}a^{m-j}b^{j}\xi^{m-j}\varpi^{\frac{j}{2}}-(b+2)\sum\limits_{2\nmid
j}{m\choose j}a^{m-j}b^{j}\xi^{m-j+1}\varpi^{\frac{j-1}{2}})\\
\end{eqnarray}
\begin{eqnarray}
\nonumber c_{r,1}&=&\frac{\sqrt{2(b+1)\eta}}{\sqrt{\varpi+(b+2)\xi
\sqrt{\varpi}}}\cdot\frac{\hat
a_{r}((b+2)\xi+\sqrt{\varpi})}{\sqrt{2\eta(\varpi+(b+2)\xi\sqrt{\varpi})}}\cdot
\frac{(a\xi+b\sqrt{\varpi})^{m}}{2^{m}(b+1)^{m}}\\\nonumber&&+
\frac{\sqrt{2(b+1)\eta}}{\sqrt{\varpi-(b+2)\xi
\sqrt{\varpi}}}\cdot\frac{\hat
a_{r}((b+2)\xi-\sqrt{\varpi})}{\sqrt{2\eta(\varpi-(b+2)\xi\sqrt{\varpi})}}\cdot
\frac{(a\xi-b\sqrt{\varpi})^{m}}{2^{m}(b+1)^{m}}\\&=& \frac{\hat
a_{r}}{2^{m-1}(b+1)^{m-\frac{1}{2}}}\sum\limits_{2\nmid j}{m\choose
j}a^{m-j}b^{j}\xi^{m-j}\varpi^{\frac{j-1}{2}}
\end{eqnarray} for $2\leq r\leq n$,
\begin{eqnarray}
\nonumber c_{r,s}&=&-(\frac{\hat
a_{r}((b+2)\xi+\sqrt{\varpi})}{\sqrt{2\eta(\varpi+(b+2)\xi\sqrt{\varpi})}}\cdot
\frac{\hat
a_{s}((b+2)\xi+\sqrt{\varpi})}{\sqrt{2\eta(\varpi+(b+2)\xi\sqrt{\varpi})}}\\\nonumber&&+
\frac{\hat
a_{r}((b+2)\xi-\sqrt{\varpi})}{\sqrt{2\eta(\varpi-(b+2)\xi\sqrt{\varpi})}}\cdot
\frac{\hat
a_{s}((b+2)\xi-\sqrt{\varpi})}{\sqrt{2\eta(\varpi-(b+2)\xi\sqrt{\varpi})}})\xi^{m}
\\\nonumber&&
+\frac{\hat
a_{r}((b+2)\xi+\sqrt{\varpi})}{\sqrt{2\eta(\varpi+(b+2)\xi\sqrt{\varpi})}}\cdot
\frac{\hat
a_{s}((b+2)\xi+\sqrt{\varpi})}{\sqrt{2\eta(\varpi+(b+2)\xi\sqrt{\varpi})}}\cdot
\frac{(a\xi+b\sqrt{\varpi})^{m}}{2^{m}(b+1)^{m}}\\\nonumber&&
+\frac{\hat
a_{r}((b+2)\xi-\sqrt{\varpi})}{\sqrt{2\eta(\varpi-(b+2)\xi\sqrt{\varpi})}}\cdot
\frac{\hat
a_{s}((b+2)\xi-\sqrt{\varpi})}{\sqrt{2\eta(\varpi-(b+2)\xi\sqrt{\varpi})}}\cdot
\frac{(a\xi-b\sqrt{\varpi})^{m}}{2^{m}(b+1)^{m}}\\\nonumber&=& -\hat
a_{1}\hat a_{r}\hat a_{s}\xi^{m-1}+\frac{\hat a_{1}\hat a_{r}\hat
a_{s}}{2^{m}(b+1)^{m}}(\sum\limits_{2\mid
j}{m\choose j}a^{m-j}b^{j}\xi^{m-j-1}\varpi^{\frac{j}{2}}\\
&&+(b+2)\sum\limits_{2\nmid j}{m\choose
j}a^{m-j}b^{j}\xi^{m-j}\varpi^{\frac{j-1}{2}})
\end{eqnarray}
for $2\leq r, s\leq n$ with $r\neq s$, and
\begin{eqnarray}
\nonumber c_{r,r}&=&\hat a_{1}(\eta-\hat a_{r}^{2})\xi^{m-1}+
\frac{\hat a_{1}\hat a_{r}^{2}}{2^{m}(b+1)^{m}}(\sum\limits_{2\mid
j}{m\choose
j}a^{m-j}b^{j}\xi^{m-j-1}\varpi^{\frac{j}{2}}\\
&&+(b+2)\sum\limits_{2\nmid j}{m\choose
j}a^{m-j}b^{j}\xi^{m-j}\varpi^{\frac{j-1}{2}})
\end{eqnarray}
for $2\leq r\leq n$.

 For convenience, we denote
\begin{equation}
f(m,s)=\sum\limits_{r=s}^{\llbracket\frac{m}{2}\rrbracket}
\frac{4^{s}(b+1)^{s}(b+2)^{2r-2s}a^{m-2r}b^{2r}}{2^{m}(b+1)^{m}}
{r\choose s}{m\choose 2r}
\end{equation} and
\begin{equation}
g(m,s)=\sum\limits_{r=s}^{\llbracket\frac{m}{2}\rrbracket}
\frac{4^{s}(b+1)^{s}(b+2)^{2r-2s}a^{m-2r-1}b^{2r+1}}{2^{m}(b+1)^{m}}
{r\choose s}{m\choose 2r+1}
\end{equation}
Thus we have
\begin{eqnarray}
(\hat a_{1}^{m}B^{m})(\hat a_1,\hat a_2,...,\hat a_n)= \left(
\begin{array}{cccc}
\hat a_{1}^{m}c_{1,1}&\frac{c_{1,2}\hat a_{1}^{m}}{\sqrt{b+1}}
&\cdots&\frac{c_{1,n}\hat a_{1}^{m}}{\sqrt{b+1}}\\
\sqrt{b+1}\hat a_{1}^{m}c_{2,1}&\hat a_{1}^{m}c_{2,2}&\cdots&\hat a_{1}^{m}c_{2,n}\\
\vdots&\vdots&\ddots&\vdots\\
\sqrt{b+1}\hat a_{1}^{m}c_{n,1}&\hat a_{1}^{m}c_{n,2}&\cdots&\hat
a_{1}^{m}c_{n,n}
\end{array}
\right)
\end{eqnarray}
for $m\geq 1$, where
\begin{equation}
\hat
a_{1}^{m}c_{1,1}=\sum\limits_{s=0}^{\llbracket\frac{m}{2}\rrbracket}(f(m,s)-(b+2)g(m,s))\hat
a_{1}^{2m-2s}\eta^{m-s},
\end{equation}
\begin{equation}
\frac{\hat
a_{1}^{m}c_{1,j}}{\sqrt{b+1}}=\sum\limits_{s=0}^{\llbracket\frac{m-1}{2}\rrbracket}
2g(m,s)\hat a_{1}^{2m-2s-1}\eta^{m-s-1}\hat
a_{j},\;\;\;\;\;\;\;\;\;\;\;\;\;\;\;\;\;
\end{equation}
\begin{equation}
\sqrt{b+1}\hat
a_{1}^{m}c_{j,1}=\sum\limits_{s=0}^{\llbracket\frac{m-1}{2}\rrbracket}
2(b+1)g(m,s)\hat a_{1}^{2m-2s-1}\eta^{m-s-1}\hat
a_{j}\;\;\;\;\;\;\;\;\;\;\;\;\;\;\;\;
\end{equation}
for $2\leq j\leq n$ and
\begin{equation}
\hat a_{1}^{m}c_{j,l}=\delta_{j,l}\hat a_{1}^{2m}\eta^{m}+
\sum\limits_{s=0}^{\llbracket\frac{m}{2}\rrbracket}
(-\delta_{s,0}+f(m,s)+(b+2)g(m,s))\hat a_{1}^{2m-2s}\eta^{m-s-1}\hat
a_{j}\hat a_{l}
\end{equation}
for $2\leq j,\  l\leq n$ by (3.73), (3.83)-(3.89). According to
(3.72),
\begin{equation}(T_{0}^{-})^{m}(T_{1}+T_{0}^{-}T_{2})^{m}=
(-1)^m(\hat a_{1}^{m}B^{m})
\left(\int_{(x_{1})},\partial_{x_2},\ldots,\partial_{x_{n}}
\right).\end{equation}

Set
\begin{equation}\Delta_{2,n}=\sum\limits_{r=2}^{n}\partial_{x_{r}}^{2}.\end{equation}
By (3.66) and (3.89)-(3.94), we have the following solutions
\begin{eqnarray}
\vec{u}_{j}(\epsilon,l_{2},\ldots,l_{n})=\left(\begin{array}{c}
u_{j}^{1}(\epsilon,l_{2},\ldots,l_{n})\\
\vdots\\u_{j}^{n}(\epsilon,l_{2},\ldots,l_{n})
\end{array}
\right)
\end{eqnarray}of
Navier equations, where
\begin{eqnarray}\nonumber
u_{1}^{1}&=&x_{1}^{\epsilon}\prod\limits_{q=2}^{n}x_{q}^{l_{q}}+
\sum\limits_{m=1}^{\infty}\sum\limits_{s=0}^{\llbracket\frac{m}{2}\rrbracket}(-1)^{m}
(f(m,s)-(b+2)g(m,s))\\
& &\times\frac{x_{1}^{\epsilon+2m-2s}}{(\epsilon+2m-2s)!}
\Delta_{2,n}^{m-s}(\prod\limits_{q=2}^{n}x_{q}^{l_{q}}),
\end{eqnarray}
\begin{eqnarray}
u_{1}^{r}=\sum\limits_{m=1}^{\infty}\sum\limits_{s=0}^{\llbracket\frac{m}{2}\rrbracket}(-1)^{m}
2(b+1)g(m,s)l_{r}\frac{x_{1}^{\epsilon+2m-2s-1}}{(\epsilon+2m-2s-1)!}
\Delta_{2,n}^{m-s-1}(x_{r}^{-1}\prod\limits_{q=2}^{n}x_{q}^{l_{q}})
\end{eqnarray}
for $2\leq r\leq n$,
\begin{eqnarray}
u_{j}^{1}=\sum\limits_{m=1}^{\infty}\sum\limits_{s=0}^{\llbracket\frac{m-1}{2}\rrbracket}(-1)^{m}
2g(m,s)l_{j}\frac{x_{1}^{\epsilon+2m-2s-1}}{(\epsilon+2m-2s-1)!}
\Delta_{2,n}^{m-s-1}(x_{j}^{-1}\prod\limits_{q=2}^{n}x_{q}^{l_{q}})
\end{eqnarray}
for $2\leq j\leq n$,
\begin{eqnarray}\nonumber
u_{j}^{r}&=&\sum\limits_{m=1}^{\infty}\sum\limits_{s=0}^{\llbracket\frac{m-1}{2}\rrbracket}(-1)^{m}
(-\delta_{s,0}+f(m,s)+(b+2)g(m,s))l_{r}l_{j}\nonumber\\&&\times
\frac{x_{1}^{\epsilon+2m-2s}}{(\epsilon+2m-2s)!}
\Delta_{2,n}^{m-s-1}(x_{r}^{-1}x_{j}^{-1}\prod\limits_{q=2}^{n}x_{q}^{l_{q}})
\end{eqnarray} for $2\leq r,\ j\leq n$ with $r\neq j$, and
\begin{eqnarray}\nonumber
u_{j}^{j}&=&x_{1}^{\epsilon}\prod\limits_{q=2}^{n}x_{q}^{l_{q}}+
\sum\limits_{m=1}^{\infty}(-1)^{m}\frac{x_{1}^{\epsilon+2m}}{(\epsilon+2m)!}
\Delta_{2,n}^{m}(\prod\limits_{q=2}^{n}x_{q}^{l_{q}})\\\nonumber&&+
\sum\limits_{m=1}^{\infty}\sum\limits_{s=0}^{\llbracket\frac{m}{2}\rrbracket}(-1)^{m}
(-\delta_{s,0}+f(m,s)+(b+2)g(m,s))l_{j}(l_{j}-1)\\&&\times
\frac{x_{1}^{\epsilon+2m-2s}}{(\epsilon+2m-2s)!}
\Delta_{2,n}^{m-s-1}(x_{j}^{-2}\prod\limits_{q=2}^{n}x_{q}^{l_{q}})
\end{eqnarray}
 for $2\leq j\leq n$.\vspace{0.4cm}

 \begin{theorem}
 The set
$\{\ \vec{u}_{j}(\epsilon,l_{2},\ldots,l_{n})\ |\ \epsilon=0\ or\
1;\ l_{r}\in\mathbb{N};\ j=1,\ldots,n\}$ forms a basis of the space
of polynomial solutions for  Navier equations. In particular,
\begin{equation}
\mbox{\it dim}\:\hat{\mathcal H}_{k}=n\cdot\mbox{\it dim}\:\mathcal
{H}_{k}.
\end{equation}
\end{theorem}

\section{Initial Value Problems}

In this section we will deal with the initial value problems of
Navier equations and Lam\'{e} equations using Lemma 3.7 and Fourier
expansions.

Recall that the initial value problem of Navier equations is as
follows:
\begin{equation} \label{eq:2}
\left\{ \begin{aligned}
         \iota_1\Delta \vec({u})+(\iota_1+\iota_2)(\nabla^{T}\cdot\nabla) (\vec{u}) & =0 \\
                   \vec{u}(0,x_{2},\ldots,x_{n}) & =
        \vec{g}_0(x_{2},\ldots,x_{n})\\
                   \vec{u}_{x_{1}}(0,x_{2},\ldots,x_{n}) & =
        \vec{g}_1(x_{2},\ldots,x_{n})
                          \end{aligned} \right.
                          \end{equation}
with $x_{1}\in\mbb{R}$  and $x_{r}\in [-a_{r},\ a_{r}]$ for
$r=2,\ldots,n$, where $a_{2},\ldots,a_{n}$ are positive numbers, and
the $j$th component $g_{\epsilon}^{j}$ of $\vec{g}_{\epsilon}$ is a
continuous function for $\epsilon=0\ or\ 1$ and $j=1,\ldots,n$. For
convenience, we denote
\begin{equation} k^\dg_r=\frac{k_r}{a_r},\;\;\vec
k^\dg=(k^\dg_2,...,k_n^\dg)^T\qquad\for\;\;\vec
k=(k_2,...,k_n)^T\in\mbb{N}^{\:n-1},\end{equation} and
\begin{equation}\vec{x}=(x_{2},\ldots,x_{n})^T,\qquad
\vec{k}^{\dag}\cdot\vec{x}=\sum\limits_{r=2}^{n}a_{r}^{-1}k_{r}x_{r}.
\end{equation}

By lemma 3.7, we have that
\begin{eqnarray}
\vec{\phi}_{\epsilon}(x_1,\ldots,x_n)=\sum\limits_{m=0}^{\infty}
(T_{0}^{-})^{m}(T_{1}+T_{0}^{-}T_{2})^{m}(x_1^{\epsilon}{\vec
g}_{\epsilon})
\end{eqnarray} and
\begin{eqnarray}
\vec{\psi}_1(x_1,\ldots,x_n)=\sum\limits_{m=0}^{\infty}
(T_{0}^{-})^{m}(T_{1}+T_{0}^{-}T_{2})^{m}(x_1\partial_{x_1}
(\vec{\phi}_0(0,x_2,\ldots,x_n)))
\end{eqnarray} are solutions of Navier equations. Denote
\begin{eqnarray}
\vec u(x_1,\ldots,x_n)=\vec{\phi}_0(x_1,\ldots,x_n)+
\vec{\phi}_1(x_1,\ldots,x_n)-\vec{\psi}_1(x_1,\ldots,x_n).
\end{eqnarray} Then by (3.90)-(3.94) and superposition
principle, the function $\vec u(x_1,\ldots,x_n)$ is the solution of
(4.1). Now we give the explicit expression of $\vec
u(x_1,\ldots,x_n)$. For convenience, we write that
\begin{eqnarray}
\vec{\phi}_{\epsilon}=\sum_{j=1}^n\phi_{\epsilon}^j\vs_j,\;\;\;\;
\vec{\psi}_1=\sum_{j=1}^n\vec{\psi}_1^j\vs_j.
\end{eqnarray}
Take the Fourier expansions of ${\vec g}_{\epsilon}$:

\begin{eqnarray}
{\vec
g}_{\epsilon}=\sum_{j=1}^n\sum_{\vec{0}\preceq\vec{k}\in\mathbb{N}^{n-1}}
(b_{\epsilon}^j(\vec k)\cos2\pi(\vec{k}^{\dag}\cdot\vec{x})+
c_{\epsilon}^j(\vec k)\sin2\pi(\vec{k}^{\dag}\cdot\vec{x}))\vs_j,
\end{eqnarray} where
\begin{eqnarray}
b_{\epsilon}^{j}(\vec{k})&=&\frac{1}{2^{n-2}a_{2}\cdots
a_{n}}\int_{-a_{2}}^{a_{2}}\cdots\int_{-a_{n}}^{a_{n}}g_{\epsilon}^{j}(x_{2},\ldots,x_{n})
\cos2\pi(\vec{k}^{\dag}\cdot\vec{x})dx_{n}\cdots dx_{2},
\end{eqnarray}
\begin{eqnarray}
c_{\epsilon}^{j}(\vec{k})&=&\frac{1}{2^{n-2}a_{2}\cdots
a_{n}}\int_{-a_{2}}^{a_{2}}\cdots\int_{-a_{n}}^{a_{n}}g_{\epsilon}^{j}(x_{2},\ldots,x_{n})
\sin2\pi(\vec{k}^{\dag}\cdot\vec{x})dx_{n}\cdots dx_{2}.
\end{eqnarray} Substituting (4.8) into (4.4), and using (3.90)-(3.94) again, we get that
\begin{eqnarray}
\phi_{\epsilon}^1&=&\sum_{\vec{0}\preceq\vec{k}\in\mathbb{N}^{n-1}}\sum_{m=0}^{\infty}
(\sum_{s=0}^{\llbracket\frac{m}{2} \rrbracket
}(-1)^s4^{m-s}(f(m,s)-(b+2)g(m,s))\frac{x_{1}^{\epsilon+2m-2s}}{(\epsilon+2m-2s)!}
\nonumber\\
&&\times(b_{\epsilon}^1(\vec k)\cos2\pi(\vec{k}^{\dag}\cdot\vec{x})+
c_{\epsilon}^1(\vec k)\sin2\pi(\vec{k}^{\dag}\cdot\vec{x}))
(\sum\limits_{r=2}^{n}(k_{r}^{\dag}\pi)^{2})^{m-s}\nonumber\\&&+\sum_{l=2}^n
\sum_{s=0}^{\llbracket\frac{m-1}{2}\rrbracket}(-1)^s4^{m-s}g(m,s)\pi
k_l^{\dag}\frac{x_{1}^{\epsilon+2m-2s-1}}{(\epsilon+2m-2s-1)!}\nonumber\\&&
\times(b_{\epsilon}^l(\vec k)\sin2\pi(\vec{k}^{\dag}\cdot\vec{x})-
c_{\epsilon}^l(\vec k)\cos2\pi(\vec{k}^{\dag}\cdot\vec{x}))
(\sum\limits_{r=2}^{n}(k_{r}^{\dag}\pi)^{2})^{m-s-1}),
\end{eqnarray}
\begin{eqnarray}
\phi_{\epsilon}^j&=&\sum_{\vec{0}\preceq\vec{k}\in\mathbb{N}^{n-1}}\sum_{m=0}^{\infty}
(\sum_{s=0}^{\llbracket\frac{m-1}{2} \rrbracket
}(-1)^s4^{m-s}(b+1)g(m,s)\pi
k_{j}^{\dag}\frac{x_{1}^{\epsilon+2m-2s-1}}{(\epsilon+2m-2s-1)!}
\nonumber\\
&&\times(b_{\epsilon}^1(\vec k)\sin2\pi(\vec{k}^{\dag}\cdot\vec{x})-
c_{\epsilon}^1(\vec k)\cos2\pi(\vec{k}^{\dag}\cdot\vec{x}))
(\sum\limits_{r=2}^{n}(k_{r}^{\dag}\pi)^{2})^{m-s-1}\nonumber\\&&
+4^m\frac{x_{1}^{\epsilon+2m}}{(\epsilon+2m)!} (b_{\epsilon}^j(\vec
k)\cos2\pi(\vec{k}^{\dag}\cdot\vec{x})+ c_{\epsilon}^j(\vec
k)\sin2\pi(\vec{k}^{\dag}\cdot\vec{x}))
(\sum\limits_{r=2}^{n}(k_{r}^{\dag}\pi)^{2})^{m}
\nonumber\\&&+\sum_{l=2}^n
\sum_{s=0}^{\llbracket\frac{m}{2}\rrbracket}(-1)^{s}4^{m-s}\pi^2k_j^{\dag}k_l^{\dag}
(-\delta_{s,0}+f(m,s)+(b+2)g(m,s))
\frac{x_{1}^{\epsilon+2m-2s}}{(\epsilon+2m-2s)!}\nonumber\\&&
\times(b_{\epsilon}^l(\vec k)\cos2\pi(\vec{k}^{\dag}\cdot\vec{x})+
c_{\epsilon}^l(\vec k)\sin2\pi(\vec{k}^{\dag}\cdot\vec{x}))
(\sum\limits_{r=2}^{n}(k_{r}^{\dag}\pi)^{2})^{m-s-1})
\end{eqnarray} for $j=2,\ldots,n,$
\begin{eqnarray}
\psi_1^1&=&\sum_{\vec{0}\preceq\vec{k}\in\mathbb{N}^{n-1}}\sum_{m=0}^{\infty}
(\sum_{l=2}^n\sum_{s=0}^{\llbracket\frac{m}{2}\rrbracket}\frac{(-1)^{s}4^{m-s}\cdot2b\pi
k_{l}^{\dag}}{b+1}(f(m,s)-(b+2)g(m,s))
\frac{x_{1}^{1+2m-2s}}{(1+2m-2s)!}\nonumber\\&&\times (b_0^l(\vec
k)\sin2\pi(\vec{k}^{\dag}\cdot\vec{x})-c_0^l(\vec
k)\cos2\pi(\vec{k}^{\dag}\cdot\vec{x}))
(\sum\limits_{r=2}^{n}(k_{r}^{\dag}\pi)^{2})^{m-s}\nonumber\\&& +
\sum_{l=2}^n\sum_{s=0}^{\llbracket\frac{m-1}{2}\rrbracket}
(-1)^{s+1}4^{m-s}\cdot2b(\pi k_{l}^{\dag})^2g(m,s)
\frac{x_{1}^{2m-2s}}{(2m-2s)!}\nonumber\\&&\times (b_0^1(\vec
k)\cos2\pi(\vec{k}^{\dag}\cdot\vec{x})+c_0^1(\vec
k)\sin2\pi(\vec{k}^{\dag}\cdot\vec{x}))
(\sum\limits_{r=2}^{n}(k_{r}^{\dag}\pi)^{2})^{m-s-1} ),
\end{eqnarray} and
\begin{eqnarray}
&&\psi_1^j=\sum_{\vec{0}\preceq\vec{k}\in\mathbb{N}^{n-1}}\sum_{m=0}^{\infty}
(\sum_{l=2}^n\sum_{s=0}^{\llbracket\frac{m-1}{2} \rrbracket
}(-1)^{s+1}4^{m-s}\cdot2b\pi^2
k_{l}^{\dag}k_{j}^{\dag}g(m,s)\frac{x_{1}^{2m-2s}}{(2m-2s)!}
\nonumber\\
&&\times(b_0^l(\vec k)\cos2\pi(\vec{k}^{\dag}\cdot\vec{x})+
c_0^l(\vec k)\sin2\pi(\vec{k}^{\dag}\cdot\vec{x}))
(\sum\limits_{r=2}^{n}(k_{r}^{\dag}\pi)^{2})^{m-s-1}\nonumber\\&&
+4^m\cdot2b\pi k_{j}^{\dag}\frac{x_{1}^{1+2m}}{(1+2m)!} (b_0^1(\vec
k)\sin2\pi(\vec{k}^{\dag}\cdot\vec{x})-c_0^1(\vec
k)\cos2\pi(\vec{k}^{\dag}\cdot\vec{x}))
(\sum\limits_{r=2}^{n}(k_{r}^{\dag}\pi)^{2})^{m}
\nonumber\\&&+\sum_{l=2}^n
\sum_{s=0}^{\llbracket\frac{m}{2}\rrbracket}(-1)^{s}4^{m-s}
\cdot2\pi^3k_j^{\dag}(k_l^{\dag})^2
(-\delta_{s,0}+f(m,s)+(b+2)g(m,s))
\frac{x_{1}^{1+2m-2s}}{(1+2m-2s)!}\nonumber\\&& \times(b_0^1(\vec
k)\sin2\pi(\vec{k}^{\dag}\cdot\vec{x})-c_0^1(\vec
k)\cos2\pi(\vec{k}^{\dag}\cdot\vec{x}))
(\sum\limits_{r=2}^{n}(k_{r}^{\dag}\pi)^{2})^{m-s-1})
\end{eqnarray} for $j=2,\ldots,n.$ Thus, as we mentioned, by superposition
principle and Fourier expansions, we get
\begin{theorem} The solution of (4.1) is
\begin{eqnarray}
\vec{u}(x_1,\ldots,x_n)=\sum_{l=1}^n(\sum_{\epsilon=0}^1\phi_{\epsilon}^l(x_1,\ldots,x_n)-
\psi_1^l(x_1,\ldots,x_n))\vs_l,
\end{eqnarray} where $\phi_{\epsilon}^l(x_1,\ldots,x_n)$ and
$\psi_1^l(x_1,\ldots,x_n)$ are defined by (4.11)-(4.14). The
convergence of the series (4.15) is guaranteed by Kovalevskaya
Theorem on the existence and uniqueness of the solution of linear
partial differential equations when the given functions in (4.1) are
analytic.
\end{theorem}

The initial value problem of Lam\'{e} equations is as follows:
\begin{equation} \label{eq:2}
\left\{ \begin{aligned}
         \vec{u}_{tt}-b^{-1}\Delta (\vec{u})-(\nabla^{T}\cdot\nabla)(\vec{u}) & =0 \\
                   \vec{u}(0,x_{1},\ldots,x_{n}) & =
        \vec{h}_{0}(x_{1},\ldots,x_{n})\\
                   \vec{u}_{t}(0,x_{1},\ldots,x_{n}) & =
        \vec{h}_{1}(x_{1},\ldots,x_{n})
                          \end{aligned} \right.
                          \end{equation}
with $x_{r}\in [-a_{r},\ a_{r}]$ for $r=1,\ldots,n$, where
$a_{1},\ldots,a_{n}$ are positive numbers and the $j$th component
$h_{\epsilon}^{j}$ of $\vec{h}_{\epsilon}$ is continuous function
for $\epsilon=0\ \mbox{or}\ 1$ and $j=1,\ldots,n$.

Set
\begin{equation}T_{1}=\partial_{t}^{2}I_{n},\qquad T_{2}=b^{-1}\Delta
I_{n}+H,\qquad \mbox{where}\;\;
H=\nabla^{T}\cdot\nabla.\end{equation} Then by Lemma 3.7, the set
\begin{eqnarray}
\{\sum\limits_{m=0}^{\infty}(T_{1}^{-})^{m}(t^{\epsilon}(T_{2}^{m})(\vec{g}))\mid
\vec{g}\in\hat{\cal A};\ \epsilon=0\ \mbox{or}\ 1\}
\end{eqnarray}
spans the polynomial solution space of Lam\'{e} equations. Note that
\begin{eqnarray}
T_{2}^{m}=b^{-m}(\Delta^{m}I_{n}+((b+1)^{m}-1)\Delta^{m-1}H)
\end{eqnarray}
for $m\geq1$ and $T_{2}^{0}=I_{n}$.

 For
convenience, we denote
\begin{equation} k^\dg_r=\frac{k_r}{a_r},\;\;\vec
k^\dg=(k^\dg_1,...,k_n^\dg)^T\qquad\for\;\;\vec
k=(k_1,...,k_n)^T\in\mbb{N}^{\:n},\end{equation} and
\begin{equation}\vec{x}=(x_1,\ldots,x_{n})^T,\qquad
\vec{k}^{\dag}\cdot\vec{x}=\sum\limits_{r=1}^{n}a_{r}^{-1}k_{r}x_{r}.
\end{equation}
Take the Fourier expansions of ${\vec h}_{\epsilon}$:

\begin{eqnarray}
{\vec
h}_{\epsilon}=\sum_{j=1}^n\sum_{\vec{0}\preceq\vec{k}\in\mathbb{N}^{n}}
(b_{\epsilon}^j(\vec k)\cos2\pi(\vec{k}^{\dag}\cdot\vec{x})+
c_{\epsilon}^j(\vec k)\sin2\pi(\vec{k}^{\dag}\cdot\vec{x}))\vs_j,
\end{eqnarray} where
\begin{eqnarray}
b_{\epsilon}^{j}(\vec{k})&=&\frac{1}{2^{n-1}a_{1}\cdots
a_{n}}\int_{-a_{1}}^{a_{1}}\cdots\int_{-a_{n}}^{a_{n}}h_{\epsilon}^{j}(x_{1},\ldots,x_{n})
\cos2\pi(\vec{k}^{\dag}\cdot\vec{x})dx_{n}\cdots dx_{1},
\end{eqnarray}
\begin{eqnarray}
c_{\epsilon}^{j}(\vec{k})&=&\frac{1}{2^{n-1}a_{1}\cdots
a_{n}}\int_{-a_{1}}^{a_{1}}\cdots\int_{-a_{n}}^{a_{n}}h_{\epsilon}^{j}(x_{1},\ldots,x_{n})
\sin2\pi(\vec{k}^{\dag}\cdot\vec{x})dx_{n}\cdots dx_{1}.
\end{eqnarray}
Note that
\begin{eqnarray}
\vec{\phi}_{\epsilon}= \sum\limits_{m=0}^{\infty}
(T_{1}^{-})^{m}(t^{\epsilon}(T_{2}^{m})(\vec{h}_{\epsilon}))
\end{eqnarray} are solutions of Lam\'{e} equations for $\epsilon=0,1$. Then by
superposition principle, the vector
\begin{eqnarray}
\vec{u}(t,x_1,\ldots,x_n)=\sum_{\epsilon=0}^1\vec{\phi}_{\epsilon}(t,x_1,\ldots,x_n)
\end{eqnarray} is also a solution. Moreover, one finds easily that
\begin{eqnarray}
\vec{u}(0,x_{1},\ldots,x_{n})=\vec{h}_{0}(x_{1},\ldots,x_{n})
\end{eqnarray} and
\begin{eqnarray}
\vec{u}_t(0,x_{1},\ldots,x_{n})=\vec{h}_{1}(x_{1},\ldots,x_{n}).
\end{eqnarray} Thus it is the solution of (4.16). Substituting
(4.22) into (4.26), we get:
\begin{theorem}
The solution of (4.16) is
\begin{eqnarray}
\vec{u}&=&\sum_{\vec{0}\preceq\vec{k}\in\mathbb{N}^{n}}\sum_{\epsilon=0}^1\sum_{j=1}^n\sum_{m=0}^{\infty}
\frac{(-1)^mt^{\epsilon+2m}}{b^m(\epsilon+2m)!}((b_{\epsilon}^j(\vec
k)\cos2\pi({\vec k}^{\dag}\cdot\vec x)+c_{\epsilon}^j(\vec
k)\sin2\pi({\vec k}^{\dag}\cdot\vec x))\nonumber\\
&&\times(\sum_{r=1}^n(2\pi
k_r^{\dag})^2)^m+\sum_{l=1}^n((b+1)^m-1)4\pi
k_j^{\dag}k_l^{\dag}\nonumber\\&&\times (b_{\epsilon}^l(\vec
k)\cos2\pi({\vec k}^{\dag}\cdot\vec x)+c_{\epsilon}^l(\vec
k)\sin2\pi({\vec k}^{\dag}\cdot\vec x))(\sum_{r=1}^n(2\pi
k_r^{\dag})^2)^{m-1})\vs_j
\end{eqnarray}
The convergence of the series (4.29) is guaranteed by Kovalevskaya
Theorem on the existence and uniqueness of the solution of linear
partial differential equations when the given functions in (4.16)
are analytic.
\end{theorem}
Set
\begin{eqnarray}
\vec{g}_r=(0,\ldots,0,\stackrel{r}{x_{1}^{l_{1}}\cdots
     x_{n}^{l_{n}}},0,\ldots,0)^T
\end{eqnarray}
and
\begin{eqnarray}
T_{2}^{m}(\vec{g}_{r})=(\tilde{g}_{1},\ldots,\tilde{g}_n)^{T}.
\end{eqnarray}
 Then by (4.19), we have
\begin{eqnarray}
\tilde{g}_{j}=\delta_{j,r}b^{-m}(\Delta^{m}(\prod\limits_{s=1}^{n}x_{s}^{l_{s}})+
l_{j}l_{r}((b+1)^{m}-1)\Delta^{m-1}
(x_{r}^{-1}x_{j}^{-1}\prod\limits_{s=1}^{n}x_{s}^{l_{s}})).
\end{eqnarray}
Thus by (4.18). we obtain

\begin{proposition}
The set
\begin{eqnarray}
&&\{t^{\epsilon}(\prod\limits_{s=1}^{n}x_{s}^{l_{s}})\vs_r
+\sum_{j=1}^n\sum_{m=1}^{\infty}\delta_{j,r}b^{-m}
\frac{t^{\epsilon+2m}}{(\epsilon+2m)!}
(\Delta^{m}(\prod\limits_{s=1}^{n}x_{s}^{l_{s}})+
l_{j}l_{r}((b+1)^{m}-1)\nonumber\\&&\times\Delta^{m-1}
(x_{r}^{-1}x_{j}^{-1}\prod\limits_{s=1}^{n}x_{s}^{l_{s}}))\vs_j \mid
\ r=1,\ldots,n,\ \ \epsilon=0\ \mbox{or}\ 1,\ l_s\in\mathbb{N} \}
\end{eqnarray}
forms a basis of polynomial solution space of Lam\'{e} equations.
\end{proposition}
\vspace{0.5cm}

 \textbf{Acknowledgement:} I would like to thank
Professor Xiaoping Xu for his advice and suggesting this research
topic.

\vspace{1cm}

\noindent{\Large \bf References}

\hspace{0.5cm}

\begin{description}

\item[{[G]}] M.Gurtin,  The linear theory of elasticity, in {\it
Handuch der Physik}, Truesdell, C., Ed., YI a/2, Springer-Verlag,
New York, 1972.

\item[{[GE]}] M.D.Gould, S.A.Edward, Enveloping Algebra Annihilators and Projection
 Techniques for Finitedimensional Cyclic Modules of a Semisimple Lie
 Algebra, {\it J. Math. Phys.} {\bf 25} (1984), no. 10,  2848-2855.

\item[{[H]}] J.Humphreys, \textit{Inroduction to Lie Algebras and Representation Theory},
Springer-Verlag, New York, 1972.

\item[{[L]}] C. Luo, Noncanonical polynomial representations of
classical Lie algebras, {\it arXiv: 0804.0305 [math.RT]}.

\item[{[M]}] L.Malvern, \textit{Inroduction to Mechanics of a Continuous Medium},
Prentice-Hall, Englewood Cliffs, NJ (1969).

\item[{[O]}] P. Olver, Conservation laws in elasticity, II, Linear
homogeneous isotropic elastostatics, {\it Arch. Rat. Mech. Anal.}
{\bf 85} (1984), 131-160.

\item[{[\"{O}]}] T.\"{O}zer, The solutions of Navier equations of classical elasticity using
 Lie symmetry groups, {\it Mech. Res. Commun.} {\bf 30} (2003), 193-201.

\item[{[R]}] A.Rodionov, Explicit solution for Lam\'{e} and other PDE systems,
{\it Appl. Math. Lett.} {\bf 51} (2006), 583-595.

\item[{[T]}] C.Truesdell, In:\textit{The linear Theory of Elasticity in the Encyclopedia
of Physics, vol.2}, Springer-verlag, Berlin (1972), 1-296.

\item[{[X1]}] X.Xu, Flag partial differential equations and representations
of Lie algebras, {\it Acta. Appl. Math.} {\bf 102} (2008), 249-280.

\item[{[X2]}] X.Xu, Matrix-Differential-Operator approach to the Maxwell equations
and the Dirac equation, {\it Acta. Appl. Math.} {\bf 102} (2008),
237-247.

\end{description}

\end{document}